\documentclass{article}

\usepackage{PRIMEarxiv}

\usepackage[utf8]{inputenc} 
\usepackage[T1]{fontenc}    
\usepackage{hyperref}       
\usepackage{url}            
\usepackage{booktabs}       
\usepackage{amsfonts}       
\usepackage{nicefrac}       
\usepackage{microtype}      
\usepackage{lipsum}
\usepackage{fancyhdr}
\usepackage{amsmath}
\usepackage{graphicx}       
\usepackage{subcaption} 
\usepackage{natbib}
\graphicspath{{media/}}     

\title{Velocity Model Building and Editing with Guided Denoising
Diffusion Implicit Models
}

\author{
  Francesco Brandolin, Tariq Alkhalifah \\
  King Abdullah University of Science and Technology (KAUST) \\
  Thuwal, Kingdom of Saudi Arabia\\
  \texttt{\{francesco.brandolin, tariq.alkhalifah\}@kaust.edu.sa} \\
}

\begin{document}
\maketitle


\begin{abstract}
Velocity-model building is a fundamental component of seismic imaging workflows, yet they remain challenging inverse problems due to limited data coverage, strong nonlinearity, and the need to incorporate heterogeneous sources of information. Such information, like wells, can have a localized influence on the velocity model. In this work, we introduce a unified framework for velocity model editing and velocity model building that combines learned diffusion priors with structurally preconditioned inverse problem formulations. A diffusion model is trained on high-resolution synthetic velocity models to learn a data-driven prior, which is then exploited through Denoising Diffusion Implicit Models inversion and guided sampling. For localized velocity model editing, we integrate the diffusion prior with a structurally preconditioned Tikhonov inversion, enabling controlled modification of target regions while preserving the global characteristics of the background model. For full velocity model building, we formulate a well-matching inverse problem augmented with imaging-based regularization and solve it using a conventional least-squares approach, the proposed Denoising Diffusion Implicit Model guided method, and Diffusion Posterior Sampling (DPS). Numerical experiments on synthetic benchmarks demonstrate that diffusion-based approaches consistently recover sharper interfaces and higher-resolution structures than classical least-squares inversion. Field data applications on the Viking Graben dataset further validate the robustness of the proposed framework for both editing and model building under realistic acquisition conditions. 
\end{abstract}


\section{Introduction}

Accurate velocity models have been a cornerstone of seismic imaging for more than five decades. Since the early development of migration methods, it has been well understood that errors in the velocity model directly translate into mispositioning, defocusing, and amplitude distortions of seismic reflections in seismic images. This dependence became even more critical with the advent of prestack depth migration and, later, full-waveform inversion (FWI), where the velocity model plays a central role in both wavefield propagation and imaging accuracy \cite{Tarantola1987, Virieux2009}. The importance of velocity model accuracy is further amplified in complex subsurface settings, such as in the presence of salt bodies or strong lateral heterogeneities, and in time-lapse (4D) seismic applications, where subtle velocity changes between repeated surveys are used to monitor reservoir evolution and production-induced effects. In this work, we address the integration of seismic-driven velocity analysis with well information for both velocity model editing and velocity model building within a unified framework. We define \emph{velocity model editing} (VME) as the localized modification of an existing velocity model, aimed at refining specific regions of interest, such as reservoir intervals, when new high-confidence information becomes available, while preserving the large-scale background structure. In contrast, \emph{velocity model building} (VMB) refers to the conventional global updating of the velocity model toward a more accurate representation of the entire subsurface, typically starting from a smooth or incomplete initial estimate. In our context, we will use VME to inject high resolution information in parts of the model where the corresponding data maybe be available in a way that allows the information to integrate seamlessly with the background model.\\
From a historical perspective, the need for accurate velocity information in seismic imaging motivated the development of early velocity analysis techniques. In 1955, a first velocity analysis method was introduced by \citeauthor{Dix1955}, who showed that stacking velocities estimated from normal moveout (NMO) on common-midpoint gathers can be related to subsurface interval velocities through a simple analytical relationship \cite{Dix1955}.
This result established the theoretical foundation for NMO correction and velocity model building from surface seismic measurements. Building on this foundation, \cite{Taner1969} introduced velocity spectra for common-midpoint data, providing a practical digital framework to estimate stacking velocities and enabling routine, data-driven velocity analysis in seismic processing. As seismic acquisition and computational capabilities improved, velocity model building evolved from local moveout-based analysis toward multi-dimensional inversion. In the mid-1980s, traveltime tomography was introduced as a framework to update laterally varying velocity models by inverting traveltime residuals measured from seismic data \cite{Bishop1985, ZeltSmith1992}. This class of methods casts velocity estimation as a global inverse problem, enabling the reconstruction of smooth background velocity models consistent with observed kinematic misfits. To stabilize the inversion and incorporate additional geological knowledge, well information and structural constraints have been progressively integrated into tomographic formulations, improving robustness and resolution in areas of limited seismic illumination \cite{Chiu1987, Clapp2004, Bakulin2010, Lipari2017}. In parallel with the development of traveltime tomography, migration velocity analysis (MVA) emerged as an image-domain approach in which velocity updates are inferred from kinematic inconsistencies in migrated images, providing a natural transition from moveout-based methods to more formal inverse-problem formulations \cite{AlYahya1988, Stork1991, Biondi2005}. Wave equation based approaches started to appear in the early 1990s, wave-equation tomography was introduced to overcome the limitations of ray-based approaches by accounting for finite-frequency wave propagation effects, enabling more accurate velocity updates in complex geological settings \cite{LuoSchuster1991, Woodward1992}. Among modern VMB techniques, FWI represents the most comprehensive framework, as it aims to recover the velocity model by matching the full seismic waveform \cite{Tarantola1984, Virieux2009}. When sufficient low-frequency content and adequate acquisition coverage are available, FWI can produce high-resolution models that significantly improve imaging. However, in many practical settings, FWI remains limited by its computational cost, sensitivity to cycle skipping, and strong dependence on the quality of the initial model \cite{Cui2020}. Additionally, imaging and velocity inversion of deep-buried reservoir can be very challenging in the presence of a complex overburden that scatters the seismic energy limiting the target illumination \cite{Guo2020}. In many settings, geophysicists aim to locally refine an existing model to better capture reservoir heterogeneity without performing a full inversion. Target-oriented FWI can refine velocity models in target areas but requires both wavefield recording in the vicinity of the target (or the redatuming of the surface data) and local forward modeling engine \cite{Li2022, Biondi2023}.\\
In the last decade, machine learning has emerged as a complementary approach to velocity model building, driven by advances in deep learning and the availability of large synthetic training datasets. Early studies primarily explored supervised learning strategies, in which neural networks are trained to directly map seismic data to velocity models \cite{Araya-Polo2018, Wang2018, Yang2019, Kazei2020, Zhicheng2021, Huang2023}. These approaches demonstrated the potential of deep learning to approximate complex nonlinear relationships at a fraction of the computational cost of conventional inversion, but their applicability is often limited by generalization issues, as it is sensitive to the training distribution. To address these limitations, subsequent work has focused on hybrid strategies that integrate machine learning within established inversion frameworks. In this context, neural networks are commonly used to parameterize the velocity model \cite{Wu2019, He2020, Sun2023} regularize the inverse problem while incorporating prior geological information \cite{Mosser2020, Sun2023FWIreg, Yang2023FWIreg, Taufik2024, Wang2024}, approximate preconditioners \cite{Alfarhan2024, Goren2024}, or learn transforms through convolutional neural networks to enhance convergence \cite{Saad2024} while the enforcement of physical consistency remains grounded in physics-based forward modeling and inverse problem formulations. Physics-informed neural networks and related approaches exemplify this trend by embedding wave-equation constraints directly into the learning process \cite{waheed2021, Taufik2024tomo}.\\
In parallel with seismic-driven velocity analysis, well information has long been used to constrain velocity model building. 
However, early approaches relied on direct interpolation or extrapolation of well-log measurements to construct high-resolution velocity such as nearest-neighbor \cite{Voronoi1908}, kriging \cite{Krige1951}, inverse distance \cite{Shepard1968}. Well log interpolation is often constrained by structural information as image-guided interpolation. From an inverse-problem perspective, well-log interpolation can be viewed as a severely underdetermined sampling problem, in which the velocity model is only known at a very limited number of spatial locations. Its direct inversion is ill-posed due to the sparsity of the sampling operator, and therefore requires regularization or preconditioning \cite{Hale2010Imageguided3I, Naeini2015, Chen2016, Karimi2017}. Structure-oriented preconditioning operators have been successfully applied also in least-squares migration, FWI and traveltime tomography \cite{ Ayeni2009, Guitton2012, Lipari2017, Gebre2025}, where they improve stability and convergence by embedding geological prior information directly into the inversion process. A key ingredient of structure-based preconditioning is the estimation of local geological orientation, typically obtained from local slope or dips estimation \cite{vanvliet1995, Fomel2002, Brandolin2024}. However, the reliability of such estimates can be limited to areas with complex geology or poor illumination, and they remain indirectly dependent on the accuracy of the background velocity model. These limitations motivate the exploration of alternative ways to encode geological realism, particularly when extending well-guided velocity model building toward more flexible updating strategies.\\
Here, we employ diffusion models to address inverse problems arising in VME. We focus on the practical scenario in which a background velocity model is already available from conventional velocity-model-building (VMB) workflows and must be locally refined as new information becomes available, such as well logs, reservoir localization, or seismic images. In this context, VME is formulated as a localized modification of an existing model, enabling efficient and targeted oriented updates without the computational cost of re-running a full inversion workflow. Velocity-model building then corresponds to applying the same diffusion-based procedure over the entire model domain.
We propose a diffusion-based framework for VME using Denoising Diffusion Implicit Models (DDIM; \cite{song2021ddim}). Unlike traditional Denoising Diffusion Probabilistic Models (DDPM; \cite{Ho2020ddpm}), which generate samples through a fully stochastic Markovian reverse process, DDIM admits a non-Markovian and potentially deterministic sampling scheme. When the reverse process is deterministic, the diffusion trajectory becomes approximately invertible, allowing an existing velocity model to be projected into the latent space of a pretrained diffusion model. This property enables direct modification of latent variables or intermediate diffusion states, followed by reverse sampling to reconstruct an edited velocity
model. In our case, we pretrain a diffusion model to learn the distribution of high-resolution velocity models. As a result, projecting an input velocity model onto the deterministic reverse trajectory of the learned diffusion model naturally injects high-resolution features into the initial estimate, providing a strong prior for generative velocity-model editing.
Relying solely on the learned data distribution, however, would be insufficient to constrain the solution, as the true subsurface model is unlikely to be perfectly represented by the training dataset. To address this limitation, we condition the DDIM sampling process by explicitly incorporating physical constraints. Specifically, we modify the structurally preconditioned Tikhonov formulation of \cite{Chen2016}, originally introduced for velocity-model building through structural preconditioning well-log interpolation, to guide the diffusion sampling. Our modification enables well-log information to be enforced within localized editing regions while preserving geological consistency at high resolution. The modified inverse problem is integrated into the diffusion process using a
measurement-guided strategy that decouples diffusion-model training from the inverse problem using the conventional strategy of decoupling the diffusion model training from the inverse problem solution (DPS; \cite{chung2023dps, Ravasi2025_measurement_guided_diffusion}). In this framework, the diffusion model learns a prior score function from a dataset of high-resolution velocity models, while the inverse-problem solution provides a model-space correction that enforces consistency with well-log and structural constraints. These corrections are injected during the diffusion sampling process, guiding the generative trajectory toward velocity models that simultaneously honor the learned prior and the available physical information. 
In the case of large editing regions or full VMB, we want to provide information also far away from the well. To achieve this, we design a novel regularization method for the previously mentioned preconditioned Tikhonov inversion based on imaging.
We introduce an image-derived sensitivity operator that links perturbations in the velocity model to measurable changes in the migrated image. This operator stabilizes the inverse problem and propagates structural information away from the well locations, complementing the structural smoothing imposed by the preconditioning. Within this formulation, we also consider the corresponding inverse problem in the absence of diffusion,  where the Tikhnov preconditioned inversion with and without imaging-based regularization are solved explicitly through a least-squares system. This allows us to directly compare the proposed diffusion-guided approach with a conventional inverse-problem solution and to assess the role of the learned diffusion prior. In addition, our approach is closely related to Diffusion Posterior Sampling (DPS), where the likelihood score of an inverse problem is combined with the prior score learned by a diffusion model. In classical DPS, the guidance term is typically derived from the gradient of the data-misfit with respect to the diffusion state. In contrast, we formulate the inverse problem explicitly in model space and inject its solution into the reverse DDIM dynamics as a direct velocity model update. For this reason, we compare our results with those obtained using the standard Diffusion Posterior Sampling (DPS) framework, highlighting the similarities and differences between the two approaches. All methods are evaluated on the Volve synthetic velocity model and on the Viking Graben dataset.\\
The main contributions of this paper can be summarized as follows:
\begin{itemize}
    \item We introduce a diffusion-based framework for localized \emph{velocity model editing} (or refinement), aimed at reducing the ill-posedness of velocity reconstruction from sparse well-log information by restricting the inversion to targeted regions of the model.
    
    \item We combine DDIM-based diffusion inversion with a structurally preconditioned least-squares inverse problem, enabling the explicit incorporation of well-log constraints together with imaging-based structural regularization during the reverse diffusion process.
    
    \item We propose a guidance formulation based on explicit LSQR velocity-model updates, establishing a connection between classical inverse-problem optimization and diffusion-based sampling while providing a more stable alternative to standard DPS guidance for this inverse-problem setting.
    
    \item We extend the proposed editing framework to full velocity model building by incorporating RTM-image guidance, allowing the diffusion prior to generate structurally consistent large-scale velocity reconstructions constrained by both geological priors and physical imaging information.
    
    \item We validate the proposed framework on synthetic and field-data experiments, demonstrating improved structural consistency, well-log propagation, and reconstruction stability when combining structurally preconditioned inversion with diffusion-based geological priors.
\end{itemize}

Overall, this work positions diffusion models not as a replacement for conventional inversion, but as a powerful complementary tool that enables flexible, efficient, and physically informed velocity model editing and building.

\section{Theory}

\subsection{Denoising Diffusion Models and DDIM Sampling}

Denoising diffusion probabilistic models (DDPMs) define a class of latent-variable
generative models that learn a data prior through a sequence of progressive
noising and denoising transformations \cite{Ho2020ddpm}.
Given a clean sample $\mathbf{v}_0 \sim p_{\mathrm{data}}(v)$, the forward diffusion
process is defined as a fixed Markov chain that gradually corrupts the data with
Gaussian noise,
\begin{equation}
\mathbf{v}_t = \sqrt{\alpha_t}\, \mathbf{v}_0 + \sqrt{1-\alpha_t}\,\boldsymbol{\epsilon},
\qquad
\boldsymbol{\epsilon} \sim \mathcal{N}(0,I),
\label{eq:forwardDDPM}
\end{equation}
where $t \in \{1,\dots,T\}$ denotes the diffusion step and
$\alpha_t = \prod_{s=1}^t (1-\beta_s)$ is the cumulative noise schedule with $\beta_s \in (0,1)$ controlling the variance of the Gaussian noise added at
step $s$.
As $t$ increases, the distribution of $\mathbf{v}_t$ converges to an isotropic
Gaussian, independently of the data distribution.
The generative objective consists of approximating the reverse-time process
$p_\theta(\mathbf{v}_{t-1} \mid \mathbf{v}_t)$, which is intractable in closed form.
DDPMs parameterize this reverse process by training a neural network
$\boldsymbol{\epsilon}_\theta(\mathbf{v}_t,t)$ to predict the injected noise at
each diffusion step.
Under this formulation, the reverse dynamics depend on an estimate of the
clean signal,
\begin{equation}
\hat{\mathbf{v}}_0(\mathbf{v}_t)
=
\frac{\mathbf{v}_t - \sqrt{1-\alpha_t}\,
\boldsymbol{\epsilon}_\theta(\mathbf{v}_t,t)}{\sqrt{\alpha_t}},
\end{equation}
which plays a central role in both sampling and conditional guidance.
Standard DDPM sampling follows a stochastic Markov chain that requires evaluating
all $T$ diffusion steps and introduces randomness at every iteration.\\
Denoising Diffusion Implicit Models (DDIM) generalize DDPM sampling by preserving
the same forward diffusion process while defining a non-Markovian reverse-time
trajectory \cite{song2021ddim}.
Importantly, DDIM does not modify the training procedure or the learned score
function, but instead provides a family of implicit reverse solvers that share
the same marginal distributions as DDPM.
Given a noisy realization $\mathbf{v}_t$, the DDIM update from step $t$ to $t-1$
is given by
\begin{equation}
\mathbf{v}_{t-1}
=
\sqrt{\alpha_{t-1}}\,\hat{\mathbf{v}}_0(\mathbf{v}_t)
+
\sqrt{1-\alpha_{t-1}-\sigma_t^2}\,
\boldsymbol{\epsilon}_\theta(\mathbf{v}_t,t)
+
\sigma_t\,\boldsymbol{\epsilon},
\label{eq:ddim_sampling}
\end{equation}
where $\boldsymbol{\epsilon} \sim \mathcal{N}(0,I)$ and $\sigma_t$ controls the
level of stochasticity in the reverse process.
Following \cite{song2021ddim}, the noise parameter is defined as
\begin{equation}
\sigma_t
=
\eta
\sqrt{\frac{1-\alpha_{t-1}}{1-\alpha_t}}
\sqrt{1-\frac{\alpha_t}{\alpha_{t-1}}},
\end{equation}
with $\eta \in [0,1]$ interpolating between deterministic ($\eta=0$) and fully
stochastic DDPM-like sampling ($\eta=1$).
A key property of DDIM is that the reverse process becomes deterministic when
$\eta = 0$, defining a unique trajectory between diffusion states.
This property enables consistent latent-space manipulations and accelerated
sampling using a reduced number of diffusion steps, while remaining faithful to
the learned data distribution.
In the following subsection, we exploit these properties to define a diffusion
inversion procedure and to incorporate physical constraints within guided DDIM
sampling.

\subsection{Diffusion inversion and Editing with DDIM}

When the noise control parameter is set to $\eta = 0$, the DDIM sampling update
becomes deterministic, as the stochastic term $\sigma_t \boldsymbol{\epsilon}$
vanishes for all diffusion steps.
In this case, the DDIM reverse process defines a unique reverse trajectory $ (\mathbf{v}_T, \mathbf{v}_{T-1}, \dots, \mathbf{v}_0)$ for a given terminal latent variable $\mathbf{v}_T$ and a fixed pretrained
network $\boldsymbol{\epsilon}_\theta$.
This determinism is a distinctive property of DDIM and is not shared by standard
DDPM sampling. The absence of stochasticity allows the DDIM sampling to be interpreted as a
reversible mapping between diffusion states. As a result, the same sampling equation (equation \ref{eq:ddim_sampling}) can be applied in the forward direction (from step $t$ to $t+1$), thereby mapping a given clean initial sample $\mathbf{v}_0$ into a corresponding (inverted) latent representation $\mathbf{v}_T$ that lies on the forward diffusion trajectory learned during training. This procedure is commonly referred to as \emph{DDIM inversion}.\\
Concretely, DDIM inversion is obtained by running the deterministic DDIM update
forward in time, starting from the clean model $\mathbf{v}_0$.
For $\eta = 0$, the forward update from step $t$ to $t+1$ reads
\begin{equation}
\mathbf{v}_{t+1}
=
\sqrt{\alpha_{t+1}}
\left(
\frac{\mathbf{v}_t - \sqrt{1-\alpha_t}\,
\boldsymbol{\epsilon}_\theta(\mathbf{v}_t,t)}
{\sqrt{\alpha_t}}
\right)
+
\sqrt{1-\alpha_{t+1}}\,
\boldsymbol{\epsilon}_\theta(\mathbf{v}_t,t),
\qquad t = 0,\dots,T-1.
\label{eq:ddim_inversion}
\end{equation}
By iterating this recursion, the pretrained diffusion model deterministically
projects the input velocity model $\mathbf{v}_0$ onto a terminal latent variable
$\mathbf{v}_T$ that is consistent with the learned DDIM prior. Thus, the prior diffusion training set will influence such projection. Samples of such phenomenon will be shared in the results section.\\
Importantly, during DDIM inversion the neural network is re-evaluated at each
newly generated state before advancing to the next diffusion step.
That is, the noise predictor $\boldsymbol{\epsilon}_\theta(\mathbf{v}_t,t)$ is
always evaluated at the current iterate $\mathbf{v}_t$ and the corresponding
diffusion time $t$. This avoids any circular dependency and ensures that the forward trajectory
remains well defined.\\
The stability of DDIM inversion relies on an implicit local consistency of the
learned noise field along the deterministic trajectory, namely
\begin{equation}
\boldsymbol{\epsilon}_\theta(\mathbf{v}_t,t)
\;\approx\;
\boldsymbol{\epsilon}_\theta(\mathbf{v}_{t+1},t+1),
\end{equation}
which holds for sufficiently small step sizes and well-trained
diffusion models. Under this assumption, the forward and reverse DDIM trajectories remain
approximately consistent, allowing inversion to be performed without introducing
additional modeling error. DDIM inversion therefore produces a latent representation $\mathbf{v}_T$ whose
deterministic reverse-time trajectory recovers the original model $\mathbf{v}_0$ under the learned prior.\\
\begin{figure}
    \centering
    \includegraphics[width=0.8\linewidth]{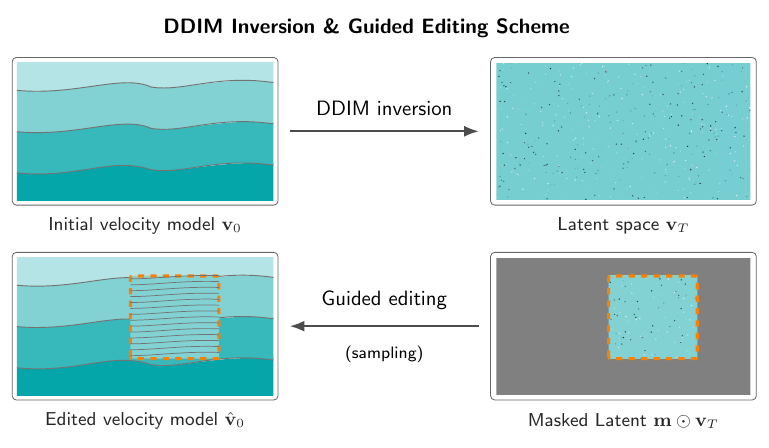}
    \caption{Overview of the DDIM-based inversion and guided editing workflow.
Top row: DDIM inversion of the initial velocity model, mapping the background model $\mathbf{v}_0$ into its latent representation $\mathbf{v}_T$.
Bottom row: Guided masked DDIM sampling, where editing is restricted to the target region (dashed orange box) through a latent-space mask and steered by data-driven constraints, while preserving the background structure outside the masked area.}
    \label{fig:DDIMinversion}
\end{figure}
We exploit this property to introduce guided DDIM sampling strategies for VME.
A schematic illustration of the diffusion inversion and guided editing sampling workflow
is shown in Figure~\ref{fig:DDIMinversion}.
The deterministic nature of DDIM inversion enables localized model editing
through simple constraints imposed during the reverse-time sampling process.
Let $\mathbf{m} \in [0,1]^{n_z n_x}$ denote a binary or smoothly varying mask
defining a region of interest.
After computing the latent representation $\mathbf{v}_T$ via DDIM inversion of the background model $\mathbf{v}_0$, we perform DDIM sampling while enforcing consistency with the \emph{inverted reference trajectory} outside the masked region.
Let $\{\mathbf{v}^{\mathrm{inv}}_t\}_{t=0}^{T}$ denote the deterministic inverted latent trajectory obtained by DDIM inversion of $\mathbf{v}_0$.
At each diffusion step $t$, this is achieved by the projection
\begin{equation}
\mathbf{v}_t \;\leftarrow\;
\mathbf{m} \odot \mathbf{v}_t + (1-\mathbf{m}) \odot \mathbf{v}^{\mathrm{inv}}_t,
\label{eq:mask_editing}
\end{equation}
where $\mathbf{v}_t$ denotes the current DDIM iterate, $\mathbf{m}$ is the binary editing mask, and $\mathbf{v}^{\mathrm{inv}}_t$ is the DDIM-inverted latent state corresponding to the background model at the same diffusion time $t$. This operation restricts generative updates to the masked region, where the effect of freshly injected Gaussian noise is retained and its magnitude is controlled by the parameter 
$\eta$. Outside the mask, the deterministic inverted latent trajectory computed via DDIM inversion is preserved in the protected region during sampling, ensuring that the background velocity structure and large-scale geological trends remain unchanged. Mask-constrained DDIM sampling therefore provides a simple and effective mechanism for localized velocity-model editing within the learned diffusion prior.\\
For VMB, instead, sampling is performed over the entire model domain and no region of the initial model needs to be preserved. In this case, the deterministic inverted trajectory is not required, and the diffusion process is warm-started from the background velocity model by injecting Gaussian noise throughout the domain, with its magnitude still controlled by the parameter 
$\eta$.\\
In both VME and VMB approaches, the reverse diffusion process is warm-started from a background velocity model rather than initialized from pure Gaussian noise. Specifically, the background model $\mathbf{v}_0$ is forward-noised to an intermediate diffusion time $t_{\mathrm{start}}$, yielding the initial latent state
\begin{equation}
\mathbf{v}_{t_{\mathrm{start}}}
=
\sqrt{\alpha_{t_{\mathrm{start}}}}\, \mathbf{v}_0
+
\sqrt{1-\alpha_{t_{\mathrm{start}}}}\, \boldsymbol{\epsilon},
\qquad
\boldsymbol{\epsilon} \sim \mathcal{N}(\boldsymbol{0}, \mathbf{I}),
\end{equation}
where $\bar{\alpha}_t$ denotes the cumulative noise schedule of the diffusion process.
In the VME case, this warm-started latent provides the initial condition for localized sampling, while hard constraints enforce consistency with the inverted background trajectory outside the editing mask, ensuring that the background velocity structure and large--scale geological trends remain unchanged.
For VMB, the same warm--start strategy is applied over the entire model domain, but no region is constrained, allowing the diffusion process to freely refine the velocity model under the learned diffusion prior and physics--based guidance.\\

\subsection{Structurally preconditioned Tikhonov inversion for velocity model updates}
Classical velocity-model building from sparse measurements, such as well logs,
is a severely ill-posed inverse problem. The data only constrain the model at a limited number of spatial locations,
making it necessary to propagate these constraints into the surrounding subsurface. Let $\mathbf{v} \in \mathbb{R}^{N}$ denote the vectorized velocity model defined on a two-dimensional grid with $N = n_z n_x$ samples, and let $\mathbf{w} \in \mathbb{R}^{N_w}$ represent the well-log measurements. The operator $ \mathbf{M} \in \mathbb{R}^{N_w \times N} $ is a restriction (sampling) operator that extracts velocity values at the well locations.
The inverse problem consists of recovering $\mathbf{v}$ from the linear system $ \mathbf{w} = \mathbf{M}\mathbf{v}. $
Since $N_w \ll N$, this system is highly underdetermined and admits infinitely many solutions. Moreover, the operator $\mathbf{M}$ provides no spatial coupling between model parameters, preventing the propagation of well-log information away from the measurement locations. A standard approach is to seek a minimum-norm solution by solving the Tikhonov-regularized least-squares problem $\min_{\mathbf{v}} \;\|\mathbf{M}\mathbf{v} - \mathbf{w}\|_2^2+\lambda^2 \|\mathbf{v}\|_2^2$, which can be equivalently written as the stacked linear system
\begin{equation}
\begin{bmatrix}
\mathbf{M} \\
\lambda\,\mathbf{I}
\end{bmatrix}
\mathbf{v}
=
\begin{bmatrix}
\mathbf{w} \\
\mathbf{0}
\end{bmatrix}.
\label{eq:unpreconditioned_stacked}
\end{equation}
While this formulation enforces stability, it promotes isotropic smoothing and
fails to honor the geological structure of the area, motivating the introduction of
structurally informed preconditioning. To regularize the problem, \cite{Chen2016} introduced a change of variables $\mathbf{v} = \mathbf{S} \mathbf{t}$, where $\mathbf{t} \in \mathbb{R}^{N}$ is a preconditioned model variable and $\mathbf{S} \in \mathbb{R}^{N \times N}$ is a structural smoothing operator that propagates information along local slopes. This reparameterization transforms the original problem into the better conditioned system $ \mathbf{w} = \mathbf{M} \mathbf{S} \mathbf{t}$,
where the operator $\mathbf{M} \mathbf{S}$ spreads the sparse well information into the model domain in a geologically consistent manner. The structural smoother $\mathbf{S} \in \mathbb{R}^{N \times N}$ is constructed as the composition $ \mathbf{S} = \mathbf{P}^{H} \mathbf{P}$,
where $ \mathbf{P} \in \mathbb{R}^{N \times N} $
is a plane-wave spray (or painting) operator aligned with local structural
slopes $\pmb{\gamma}(x,z)$, where $(x,z)$ denote the depth and horizontal spatial coordinates on the
$n_x \times n_z$ model grid.
In forward mode, $\mathbf{P}$ propagates each input sample along the estimated slopes,
spraying its value over neighboring grid points with an exponential decay away from the well.
In adjoint mode, $\mathbf{P}^{H}$ gathers contributions back along the same slope
trajectories.
The resulting operator $\mathbf{S} = \mathbf{P}^{H} \mathbf{P}$ is symmetric and positive semi-definite,
and acts as a correlation-like smoother that preferentially enforces continuity
along structural directions while avoiding smoothing across reflectors. This structurally preconditioned formulation provides a physically meaningful
way to interpolate sparse well-log information, yielding velocity models that
honor both the measurements and the underlying geological structure.
In the following, we exploit this inverse problem as the data-consistency term
used to guide DDIM sampling, allowing structural information and well
constraints to be incorporated within the diffusion-based framework.\\
Specifically, instead of directly solving for the full velocity model $\mathbf{v}$, we seek a model perturbation $\Delta \mathbf{v}$ such that the updated model $\mathbf{v}^\ast = \mathbf{v} + \Delta \mathbf{v}$ admits the structural parametrization $\mathbf{v}^\ast = \mathbf{S}(\mathbf{t} + \Delta \mathbf{t})$.
This reformulation provides a convenient way to guide the diffusion sampling. The velocity update $\Delta \mathbf{v}$ honors available well-log constraints while remaining compatible with the large-scale structural trends of the subsurface. Using $\mathbf{v} = \mathbf{S}\mathbf{t}$ and $\Delta \mathbf{v} = \mathbf{S}\Delta \mathbf{t}$, the problem can be cast as the following Tikhonov-regularized least-squares objective:
$
\min_{\Delta \mathbf{t}}
\;
\big\|
\mathbf{M}\mathbf{S}(\mathbf{t} + \Delta \mathbf{t}) - \mathbf{w}
\big\|_2^2
\;+\;
\lambda^2 \,
\|
\mathbf{t} + \Delta \mathbf{t}
\|_2^2 .
$
Defining the well-log residual as $ \mathbf{r}_w := \mathbf{w} - \mathbf{M}\mathbf{S}\mathbf{t}
= \mathbf{w} - \mathbf{M}\mathbf{v}, $
the objective can be rewritten as
\begin{equation}
\min_{\Delta \mathbf{t}}
\;
\big\|
\mathbf{M}\mathbf{S}\,\Delta \mathbf{t} - \mathbf{r}_w
\big\|_2^2
\;+\;
\lambda^2 \,
\|
\mathbf{t} + \Delta \mathbf{t}
\|_2^2 .
\label{eq:ls}
\end{equation}
This problem is equivalently expressed as the stacked linear system
\begin{equation}
\begin{bmatrix}
\mathbf{M}\mathbf{S} \\
\lambda\,\mathbf{I}
\end{bmatrix}
\Delta \mathbf{t}
=
\begin{bmatrix}
\mathbf{r}_w \\
-\lambda\,\mathbf{t}
\end{bmatrix},
\label{eq:stacked_ls}
\end{equation}
where $\lambda$ controls the strength of the regularization in the
preconditioned domain.
This system is better conditioned than the original formulation and can be
efficiently solved using conjugate-gradient or LSQR methods.
Taking the gradient of Eq. \ref{eq:ls} with respect to $\Delta \mathbf{t}$ and
setting it to zero yields the normal equations
\begin{equation}
\left(
\mathbf{S}^{\mathsf H}\mathbf{M}^{\mathsf H}\mathbf{M}\mathbf{S}
+
\lambda^2 \mathbf{I}
\right)
\Delta \mathbf{t}
=
\mathbf{S}^{\mathsf H}\mathbf{M}^{\mathsf H}
(\mathbf{w} - \mathbf{M}\mathbf{v})
-
\lambda^2 \mathbf{t}.
\end{equation}
Substituting $\mathbf{v} = \hat{\mathbf{v}}_0(\mathbf{v}_t),$ i.e., the DDIM-predicted clean sample, into the right-hand side gives the update associated with the diffusion state $\mathbf{v}_t$,
\begin{equation}
\Delta \mathbf{v}_t
=
\mathbf{S}\,\Delta \mathbf{t}
=
\mathbf{S}
\left(
\mathbf{S}^{\mathsf H}\mathbf{M}^{\mathsf H}\mathbf{M}\mathbf{S}
+
\lambda^2 \mathbf{I}
\right)^{-1}
\left[
\mathbf{S}^{\mathsf H}\mathbf{M}^{\mathsf H}
(\mathbf{w} - \mathbf{M}\hat{\mathbf{v}}_0(\mathbf{v}_t))
-
\lambda^2 \mathbf{t}
\right].
\end{equation}
This quantity represents a Hessian-preconditioned, structurally shaped update
that enforces well-log constraints while respecting the geometry of the
subsurface model.
During the final portion of the DDIM reverse trajectory, where data constraints
are allowed to influence the generative process, the standard DDIM reverse step
(Eq.~\ref{eq:ddim_sampling}) is guided using the correction
$\Delta \mathbf{v}_t$.
The resulting sampling formula reads
\begin{equation}
\mathbf{v}_{t-1}
=
\sqrt{\alpha_{t-1}}\,\hat{\mathbf{v}}_0(\mathbf{v}_t)
+
\sqrt{1-\alpha_{t-1}-\sigma_t^2}\,
\boldsymbol{\epsilon}_\theta(\mathbf{v}_t,t)
+
\sigma_t\,\boldsymbol{\epsilon}
+
\mu\,\Delta \mathbf{v}_t,
\end{equation}
where $\hat{\mathbf{v}}_0(\mathbf{v}_t)$ and
$\boldsymbol{\epsilon}_\theta(\mathbf{v}_t,t)$ are predicted by the U-Net,
$\Delta \mathbf{v}_t$ is the Hessian-preconditioned posterior update, and $\mu$
controls the update strength. 
This strategy preserves the high-resolution statistics of the diffusion prior
while ensuring consistency with the well-log constraints and the imposed
structural regularization.

\subsection{Imaging based regularization}
To inject structural information away from the well locations, we incorporate
an imaging-based regularizer.
While the smoothing operator $\mathbf{S}$ enforces geological consistency in the
vicinity of the wells, it cannot propagate constraints into regions where no
well information is available.
To address this limitation, we introduce an image-derived sensitivity operator
that links perturbations in the velocity model to measurable changes in the
migrated image. The motivation for this choice stems from the observation that migrated seismic images primarily highlight subsurface reflectors, which are associated with spatial contrasts in the underlying velocity field. Therefore, rather than directly imposing the RTM amplitudes on the reconstructed model, we seek to enforce consistency between the reflector geometry observed in the migrated image and the velocity contrasts predicted by the reconstructed model. This provides a physically motivated structural constraint that complements the sparse well information and helps guide the inversion in poorly constrained regions.\\
The scattering potential of the acoustic wave equation is given by
$q(v) = \frac{1}{v^{2}}$.
Within the linearized Born approximation, image perturbations are approximately
proportional to vertical derivatives of the scattering-contrast term
\begin{equation}
q(v) - q(v_0) = \frac{1}{v^{2}} - \frac{1}{v_0^{2}},
\end{equation}
where $v_0$ denotes a fixed background velocity.
This approximation underlies classical reverse-time migration (RTM) theory
\cite{Bleistein2001}.
Motivated by this observation, we define the nonlinear imaging forward operator
\begin{equation}
R(\mathbf v) = D_z\!\left(q(\mathbf v) - q(\mathbf v_0)\right),
\end{equation}
where $D_z:\mathbb{R}^{N} \rightarrow \mathbb{R}^{N}$ denotes a finite-difference vertical derivative operator applied
pointwise in space.\\
Let $\mathbf v$ denote the current velocity model and
$\mathbf v^\ast = \mathbf v + \Delta \mathbf v$ an updated model.
To obtain a linearized approximation of the imaging operator, we expand
$q(\mathbf v)$ to first order around $\mathbf v$.
Setting $\Delta \mathbf v = \mathbf v^\ast - \mathbf v$, the Taylor expansion
yields
\begin{equation}
q(\mathbf v^\ast)
=
q(\mathbf v + \Delta \mathbf v)
\;\approx\;
q(\mathbf v) + q'(\mathbf v)\,\Delta \mathbf v,
\qquad
q'(\mathbf v) = - \frac{2}{\mathbf v^{3}}.
\end{equation}
Applying $D_z$ and using its linearity gives the Gauss--Newton linearization
\begin{equation}
R(\mathbf v^\ast)
\;\approx\;
R(\mathbf v) + D_z\!\left[q'(\mathbf v)\,\Delta \mathbf v\right].
\end{equation}
We thus define the imaging Jacobian $\mathbf J(\mathbf v)$ through its action on
a model perturbation as
\begin{equation}
\mathbf J(\mathbf v)\,\Delta \mathbf v
\;:=\;
D_z\!\left[q'(\mathbf v)\,\Delta \mathbf v\right],
\end{equation}
so that the linearized forward map takes the standard Gauss--Newton form
\begin{equation}
R(\mathbf v^\ast) \;\approx\; R(\mathbf v) + \mathbf J(\mathbf v)\,\Delta \mathbf v .
\end{equation}
To use this linearized relation as a regularization term in the inverse problem,
we introduce the image-domain residual
\begin{equation}
\mathbf r_{\mathrm{im}}(\mathbf v)
\;=\;
\mathbf I_{\mathrm{RTM}} - R(\mathbf v),
\label{eq: imaging residual}
\end{equation}
where $\mathbf I_{\mathrm{RTM}}$ denotes the observed (fixed) migrated image obtained from the initial velocity $\mathbf v_{0}$ and
$R(\mathbf v)$ is the image predicted by the current velocity model.
For the updated model $\mathbf v^\ast = \mathbf v + \Delta \mathbf v$, substituting
the linearized forward operator into the residual definition yields
\begin{align}
\mathbf r_{\mathrm{im}}(\mathbf v^\ast)
&= \mathbf I_{\mathrm{RTM}} - R(\mathbf v^\ast) \nonumber\\
&\approx \mathbf I_{\mathrm{RTM}} -
\left( R(\mathbf v) + \mathbf J(\mathbf v)\,\Delta \mathbf v \right) \nonumber\\
&\approx \mathbf r_{\mathrm{im}}(\mathbf v) -
\mathbf J(\mathbf v)\,\Delta \mathbf v .
\end{align}
Building on the linearized image-domain residual
\(
\mathbf r_{\mathrm{im}}(\mathbf v^\ast)
\approx
\mathbf r_{\mathrm{im}}(\mathbf v) - \mathbf J(\mathbf v)\,\Delta \mathbf v
\),
we evaluate the residual at the current iterate using equation \ref{eq: imaging residual}
and seek an update parameterized through the structural preconditioner
\(
\Delta \mathbf v = \mathbf S\,\Delta \mathbf t
\).
This leads to the following combined Tikhonov objective:
\begin{equation}
\min_{\Delta \mathbf t}\;
\|\mathbf M\mathbf S\,\Delta \mathbf t - \mathbf r_w\|_2^2
\;+\;
\kappa^2
\|\mathbf J(\mathbf v)\,\mathbf S\,\Delta \mathbf t
- \mathbf r_{\mathrm{im}}\|_2^2
\;+\;
\lambda^2\,\|\mathbf t + \Delta \mathbf t\|_2^2 .
\label{eq:obj_function_imreg}
\end{equation}
This problem is equivalently expressed as the following stacked linear
least-squares system:
\begin{equation}
\begin{bmatrix}
\mathbf M \mathbf S \\
\kappa\,\mathbf J(\mathbf v)\mathbf S \\
\lambda\,\mathbf I
\end{bmatrix}
\Delta \mathbf t
=
\begin{bmatrix}
\mathbf r_w \\
\kappa\,\mathbf r_{\mathrm{im}} \\
-\lambda\,\mathbf t
\end{bmatrix}.
\end{equation}
The first term enforces consistency with the well-log measurements.
The parameter $\kappa$ controls the contribution of the imaging term, which
penalizes discrepancies between the linearized image prediction
\(
R(\mathbf v) + \mathbf J(\mathbf v)\,\mathbf S\,\Delta \mathbf t
\)
and the observed RTM image $\mathbf I_{\mathrm{RTM}}$.
The final term acts as a Tikhonov regularization in the preconditioned domain.
Together, these terms yield geologically consistent perturbations
\(
\Delta \mathbf v = \mathbf S\,\Delta \mathbf t
\),
which can be injected as guidance into the DDIM sampling stage.
Taking the gradient of Eq.~\ref{eq:obj_function_imreg} with respect to
$\Delta \mathbf t$ and setting it to zero yields the normal equations
\begin{equation}
\Big(
\mathbf S^{\mathsf H}\mathbf M^{\mathsf H}\mathbf M\mathbf S
+
\kappa^2\,\mathbf S^{\mathsf H}\mathbf J(\mathbf v)^{\mathsf H}
\mathbf J(\mathbf v)\mathbf S
+
\lambda^2\,\mathbf I
\Big)\Delta \mathbf t
=
\mathbf S^{\mathsf H}\mathbf M^{\mathsf H}
(\mathbf w - \mathbf M\mathbf S\mathbf t)
+
\kappa^2\,\mathbf S^{\mathsf H}\mathbf J(\mathbf v)^{\mathsf H}
\big(\mathbf I_{\mathrm{RTM}} - R(\mathbf v)\big)
-
\lambda^2\,\mathbf t .
\end{equation}
Given the current estimate
\(
\mathbf v = \hat{\mathbf v}_0(\mathbf v_t) = \mathbf S\,\mathbf t
\),
the Gauss--Newton model correction is
\(
\Delta \mathbf v_t = \mathbf S\,\Delta \mathbf t
\).
This update is computed efficiently using a small number of LSQR iterations,
which approximate the action of the damped inverse Hessian. During the final part of the DDIM reverse trajectory, the guided sampler updates
the velocity as
\begin{equation}
\mathbf{v}_{t-1}
=
\sqrt{\alpha_{t-1}}\,\hat{\mathbf{v}}_0(\mathbf{v}_t)
+
\sqrt{1-\alpha_{t-1}-\sigma_t^2}\,
\boldsymbol{\epsilon}_\theta(\mathbf{v}_t,t)
+
\sigma_t\,\boldsymbol{\epsilon}
+
\mu\,\Delta \mathbf v_t ,
\end{equation}
where $\mu$ denotes the update weight.
This strategy preserves the high-frequency statistics of the diffusion prior
while enforcing consistency with well-log constraints and image-derived
structural regularization.

\subsection{Diffusion Posterior Sampling}

This subsection is included to provide a baseline for comparison with the diffusion-guided inversion strategy proposed in this work. Specifically, we solve the same inverse problem described above using both our proposed diffusion framework and Diffusion Posterior Sampling (DPS). This inverse problem is addressed here using DPS for the first time.\\
DPS guides the reverse diffusion trajectory through a first-order likelihood-gradient correction. In this work, for the velocity-model editing (VME) application, DPS is implemented in combination with DDIM inversion and deterministic DDIM sampling. The initial velocity model is first projected into the diffusion latent space using DDIM inversion, after which the inverted latent trajectory is guided using the DPS update. This choice ensures a consistent and fair comparison with the proposed diffusion-guided method, as both approaches operate on the same latent representation. For velocity-model building (VMB), instead, the diffusion process is warm-started directly from the initial velocity model, following the same initialization
strategy adopted in the proposed diffusion-guided framework.\\ 
Diffusion models are trained to approximate the prior score
$\nabla_{\mathbf{v}_t}\log p(\mathbf{v}_t)$, which drives the reverse diffusion
process from noise to data according to the learned distribution of velocity
models.
When conditioning on observed well-log measurements $\mathbf{w}$, the goal is to
sample from the posterior distribution
$
p(\mathbf{v}_t \mid \mathbf{w}) \propto p(\mathbf{w} \mid \mathbf{v}_t)\,p(\mathbf{v}_t).
$
By Bayes’ rule, the posterior score decomposes as
\begin{equation}
\nabla_{\mathbf{v}_t}\log p(\mathbf{v}_t \mid \mathbf{w})
=
\nabla_{\mathbf{v}_t}\log p(\mathbf{v}_t)
+
\nabla_{\mathbf{v}_t}\log p(\mathbf{w} \mid \mathbf{v}_t),
\end{equation}
where the second term introduces a data-consistency correction that steers the
diffusion trajectory toward models consistent with the observations. Assuming a Gaussian likelihood for the well-log data $p(\mathbf{w}\mid \mathbf{v}_t)
\propto
\exp\!\left(
-\frac{
(\mathbf{M}\hat{\mathbf{v}}_0(\mathbf{v}_t)-\mathbf{w})^{\mathsf T}
(\mathbf{M}\hat{\mathbf{v}}_0(\mathbf{v}_t)-\mathbf{w})
}{2\sigma_w^2}
\right)$, evaluated at the denoised estimate $\hat{\mathbf{v}}_0(\mathbf{v}_t)$.
The exact likelihood score follows from the chain rule as
\begin{equation}
\nabla_{\mathbf{v}_t}\log p(\mathbf{w}\mid \mathbf{v}_t)
=
\frac{1}{\sigma_w^2}
\left(
\nabla_{\mathbf{v}_t}\hat{\mathbf{v}}_0(\mathbf{v}_t)
\right)^{\mathsf T}
\mathbf{M}^{\mathsf H}
\big(
\mathbf{w}-\mathbf{M}\hat{\mathbf{v}}_0(\mathbf{v}_t)
\big),
\label{eq:dps_exact_likelihood_score}
\end{equation}
where $\sigma_w^2$ denotes the measurement noise variance and
$\mathbf{M}^{\mathsf H}$ is the adjoint of the restriction operator.
Accordingly, we define the DPS descent direction as
\begin{equation}
\mathbf{g}_t
:=
-\nabla_{\mathbf{v}_t}\log p(\mathbf{w}\mid \mathbf{v}_t)
\approx
-\frac{1}{\sigma_w^2}\,
\left(
\nabla_{\mathbf{v}_t}\hat{\mathbf{v}}_0(\mathbf{v}_t)
\right)^{\mathsf T}\,
\mathbf{M}^{\mathsf H}
\big(
\mathbf{w}-\mathbf{M}\hat{\mathbf{v}}_0(\mathbf{v}_t)
\big),
\label{eq:dps_direction}
\end{equation}
which corresponds to a first-order (steepest-descent) data-consistency update in
model space. The guided DDIM update becomes
\begin{equation}
\mathbf{v}_{t-1}
=
\sqrt{\alpha_{t-1}}\,\hat{\mathbf{v}}_0(\mathbf{v}_t)
+
\sqrt{1-\alpha_{t-1}-\sigma_t^2}\,
\boldsymbol{\epsilon}_\theta(\mathbf{v}_t,t)
+
\sigma_t\,\boldsymbol{\epsilon}
-
\mu\,\mathbf{g}_t,
\label{eq:ddim_sampling_DPS}
\end{equation}
where $\hat{\mathbf{v}}_0(\mathbf{v}_t)$ and
$\boldsymbol{\epsilon}_\theta(\mathbf{v}_t,t)$ are predicted by the U-Net and
$\mu$ controls the guidance strength.
Compared to DPS, the approach proposed in this work replaces the adjoint-based
first-order update with a structurally preconditioned, Gauss--Newton-type
correction.
By explicitly incorporating geological smoothing and imaging-based sensitivity
operators, our method performs a damped, Hessian-preconditioned inversion step
within the diffusion process, which is particularly beneficial in geophysical
settings where sparse measurements make first-order guidance insufficient to
propagate constraints in a geologically consistent manner.

\section{Results}

This section evaluates the proposed diffusion-based velocity-model editing framework on both synthetic and field data. We begin with controlled \emph{synthetic experiments}, where quantitative metrics are used to assess both image-domain and model-domain performance. Structural consistency in the velocity models is evaluated using the Structural Similarity Index Measure (SSIM, \cite{Wang2004}).
Velocity accuracy along two different velocity models extracted profiles are quantified using the mean squared error (MSE),
$
\mathrm{MSE} =
\frac{1}{N} \sum_{i=1}^{N} \left( v_i - v^{\mathrm{ref}}_i \right)^2,
$
and the Pearson correlation coefficient (R-score),
$
R =
\frac{\sum_{i=1}^{N} (v_i - \bar{v})(v^{\mathrm{ref}}_i - \bar{v}^{\mathrm{ref}})}
     {\sqrt{\sum_{i=1}^{N} (v_i - \bar{v})^2}
      \sqrt{\sum_{i=1}^{N} (v^{\mathrm{ref}}_i - \bar{v}^{\mathrm{ref}})^2}},
$
where $v_i$ and $v_i^{\mathrm{ref}}$ denote, respectively, the estimated and reference velocities sampled along the well, $\bar{v}$ and $\bar{v}^{\mathrm{ref}}$ are their mean values, and $N$ is the number of depth samples.
We then apply the proposed method to \emph{field data}, focusing on Line~12 of the Viking Graben dataset, where no ground-truth velocity model is available and the evaluation is therefore based on geological plausibility, well-log consistency, and improvements in image focusing.
Finally, we present an \emph{ablation study} on the synthetic benchmark, in which the slope field used for structural guidance is computed directly from the true velocity model rather than from the RTM image. This experiment isolates the impact of slope-estimation accuracy on the diffusion-guided editing process and clarifies the role of imaging-derived structural information in the overall performance of the method.
All experiments were carried out on an Intel(R) Xeon(R) CPU @ 2.10\,GHz equipped with a single NVIDIA GeForce RTX~3090 GPU for inference and inversion, and an NVIDIA V100 GPU for diffusion model training.

\subsection{Numerical Examples on Volve synthetic model}
We first evaluate the proposed diffusion-based framework on a synthetic benchmark, considering both localized velocity-model editing and full velocity model building (VMB). A Denoising Diffusion Probabilistic Model (DDPM) with a UNet backbone is pretrained to learn a prior over high-resolution velocity models by approximating the score function $\nabla_{x_t}\log p(x_t)$, which represents the gradient of the log-prior distribution. Training is performed using 5000 high-resolution velocity models drawn from the SEAM Arid, SEAM Arid Barrett, SEAM Phase~I, Otway, and Volve velocity models, which have been normalized between -1 and 1. Importantly, the true Volve synthetic velocity model used in the subsequent tests is not included in the training set; only augmented Volve-derived models are used during training.
The training models are extracted as $256\times512$ patches and augmented through random flips, randomized background trends, and random spatial cropping followed by resampling to the target grid. The inclusion of augmented Volve-like models in the training dataset reflects a realistic industrial setting, where prior geological knowledge, regional information, and available well-log measurements can be leveraged to tailor the generation of training data toward velocity models that are representative of the specific subsurface setting of interest. Training is carried out for 50 epochs using 1000 diffusion steps, for a total of 24 hours and 9 minutes.\\
\begin{figure}
    \centering
    \includegraphics[width=1.\linewidth]{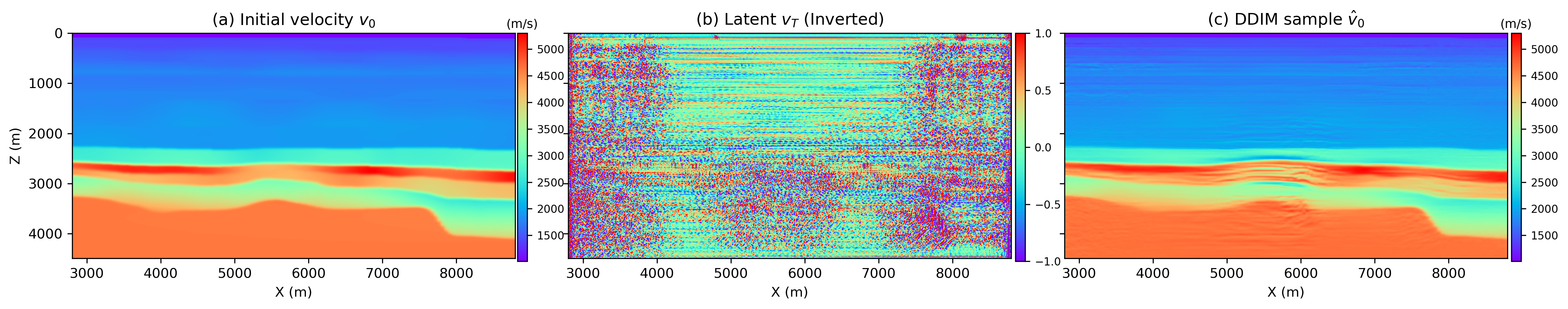}
    \caption{(a) Background velocity model $\mathbf{v}_0$.
(b) Corresponding latent representation $\mathbf{v}_T$ obtained through DDIM inversion.
(c) Reconstructed velocity model $\hat{\mathbf{v}}_0$ recovered by deterministic DDIM sampling.}
    \label{fig:synth_ddiminversion}
\end{figure}
\begin{figure}
    \centering
    \includegraphics[width=1.\linewidth]{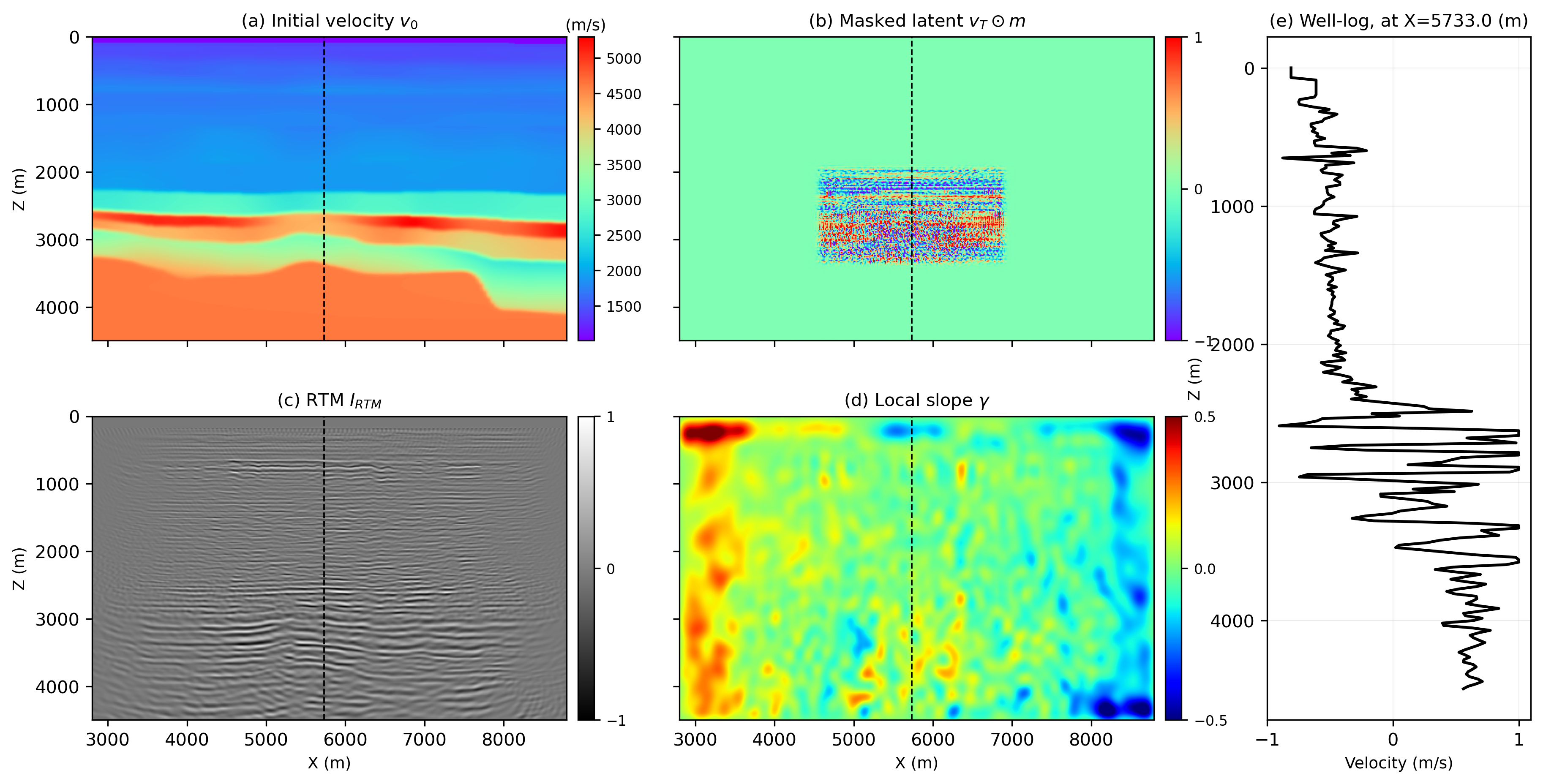}
    \caption{Synthetic test setup and velocity-profile comparison.
(a) Initial velocity model $\mathbf{v}_0$, where the black dashed line marks the well position.
(b) Masked latent representation $\mathbf{v}_T$ obtained through DDIM inversion.
(c) Reverse Time Migration (RTM) image used to derive structural guidance.
(d) Local slope field $\boldsymbol{\gamma}$ estimated from the RTM image.
(e) Well-log positioned at $x = 5733\mathrm{m}$.}
    \label{fig:synth_intro}
\end{figure}
To assess the quality of the learned diffusion prior, we first evaluate the behavior of the trained model through DDIM inversion and subsequent sampling, as illustrated in Fig.~2. Starting from the background velocity model $\mathbf{v}_0$ (Figure~\ref{fig:synth_ddiminversion}a), the model is first normalized to the range $[-1,1]$ to ensure consistency with the training data distribution, which was similarly normalized. The normalized model is then inverted into its corresponding latent representation $\mathbf{v}_T$ using DDIM inversion with 1000 timesteps (Figure~\ref{fig:synth_ddiminversion}b). Although the diffusion prior is trained by diffusing Gaussian noise, the latent representation obtained through DDIM inversion does not resemble pure Gaussian noise, as it corresponds to a conditioned and structured latent code that preserves information about the input velocity model along a valid reverse-time trajectory and is affected by the natural biases imposed by the training distribution. The inverted latent state is subsequently propagated through the deterministic DDIM reverse process to reconstruct a velocity model $\hat{\mathbf{v}}_0$ (Figure~\ref{fig:synth_ddiminversion}c). Since no data-driven or physical guidance is applied in this experiment, the reconstruction quality of $\hat{\mathbf{v}}_0$ provides a direct evaluation of the learned prior. In particular, the emergence of sharper interfaces and higher-resolution features in $\hat{\mathbf{v}}_0$ compared to the input background model reflects the high resolution nature of the training set, allowing the diffusion generation process to enrich the model with high-resolution features when projected into and sampled from the latent space. Diffusion inversion and sampling each take approximately 21 seconds on a GPU.\\
Starting from the background velocity model $\mathbf{v}_0$ (Figure~\ref{fig:synth_intro}a), where the black dashed line marks the well location, we generate synthetic seismic data by forward modeling and subsequently applying a Reverse Time Migration (RTM) image (Figure~\ref{fig:synth_intro}c). The synthetic dataset comprises 110 shot gathers, each with 4.7~seconds of recording time and 180 receivers, using temporal and receiver sampling of 2~ms and 25~m, respectively. The resulting RTM image is then used to derive structural guidance for the inversion. Local slopes $\pmb{\gamma}$ (Figure~\ref{fig:synth_intro}d) are estimated from the RTM image using plane-wave destruction \cite{Fomel2002} and are used to construct the preconditioning structural operator $\mathbf{S}$ that propagates the well information into the surrounding region.
In this synthetic example, the well-log measurement consists of a velocity profile extracted directly from the ground-truth model at the location $x = 5733\mathrm{m}$ (Figure~\ref{fig:synth_intro}e).
The initial migration velocity model $\mathbf{v}_0$ is first projected into the latent space $\mathbf{v}_T$ through DDIM inversion with $T = 600$ steps, subsampled from the full training diffusion schedule requiring approximately 13 seconds. A spatial mask $\mathbf{m}$ is then applied in the latent space following  Eq.~\ref{eq:mask_editing} to restrict the editing to the target region, yielding a masked latent representation consistent with the learned reverse-time trajectory (Figure~\ref{fig:synth_intro}b).\\
\begin{figure}
    \centering
    \includegraphics[width=1.\linewidth]{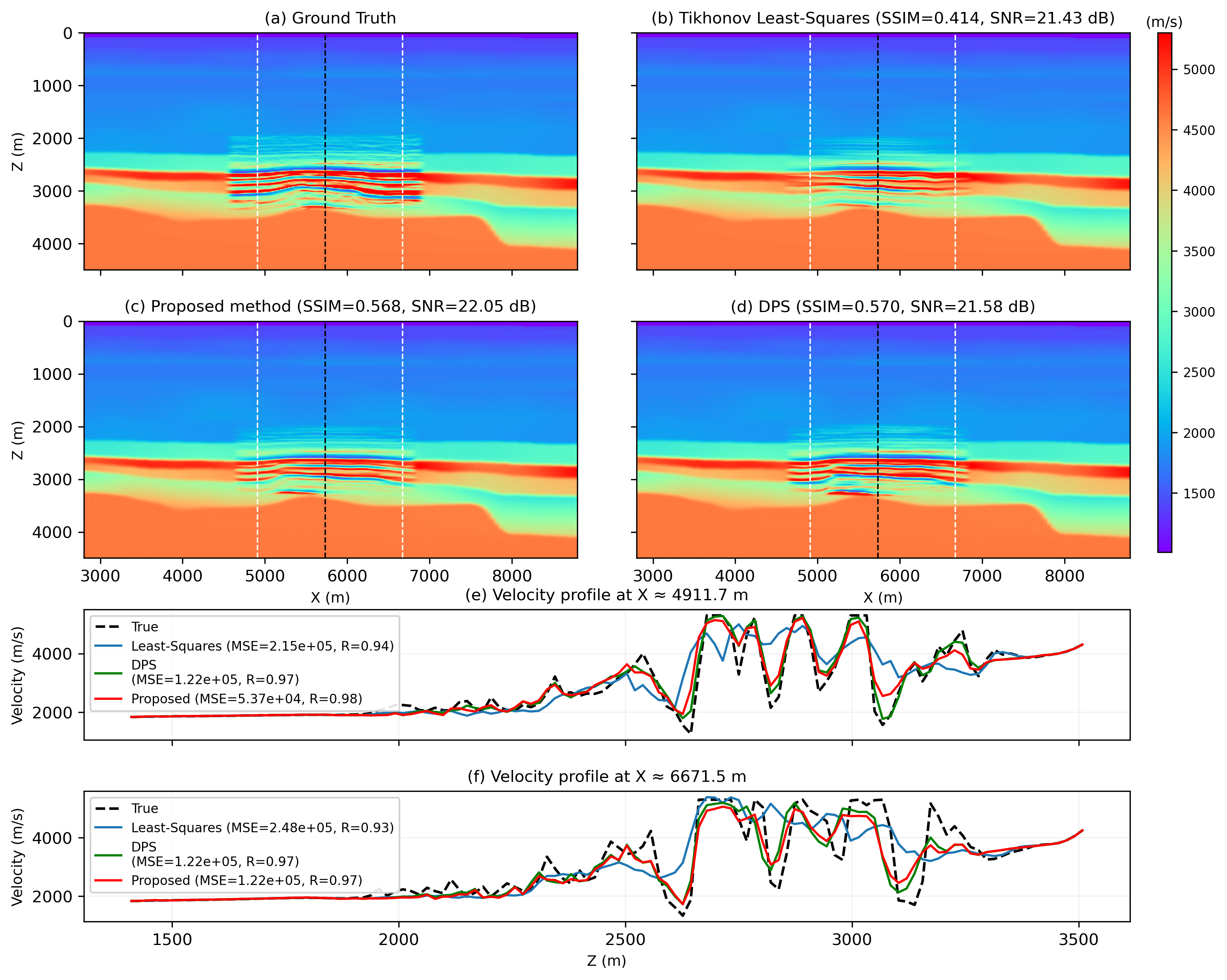}
    \caption{Results comparison for the VME synthetic test.
(a) Ground-truth velocity model.
(b) Conventional Tikhonov-regularized least-squares inversion.
(c) Velocity-model editing result obtained with the proposed diffusion-guided approach.
(d) Editing result obtained using the Diffusion Posterior Sampling (DPS) framework.
The two white dashed lines indicate the locations at which velocity profiles are extracted; the black dashed line marks the well position.
In panels (e) and (f), the black dashed curve denotes the ground-truth velocity profile, the blue curve corresponds to the Tikhonov least-squares result, the green curve to the DPS result, and the red curve to the proposed method.}
    \label{fig:synth_results}
\end{figure}
Guided editing is then performed using stochastic DDIM sampling with noise level $\eta = 0.5$. At each diffusion timestep, the update $\Delta \mathbf{v}_t$ is obtained by solving the Tikhonov-regularized system in Eq.~\ref{eq:ls} using two LSQR iterations and damping parameter $\lambda = 0.01$. The update $\Delta \mathbf{v}_t$ is stored and used as the initial guess for the LSQR solver at the subsequent diffusion step. Guidance is activated only during the final 50\% of the reverse diffusion, and the update is scaled by a factor $\mu = 0.5$.
\begin{figure}[h!]
    \centering
    \includegraphics[width=1.\linewidth]{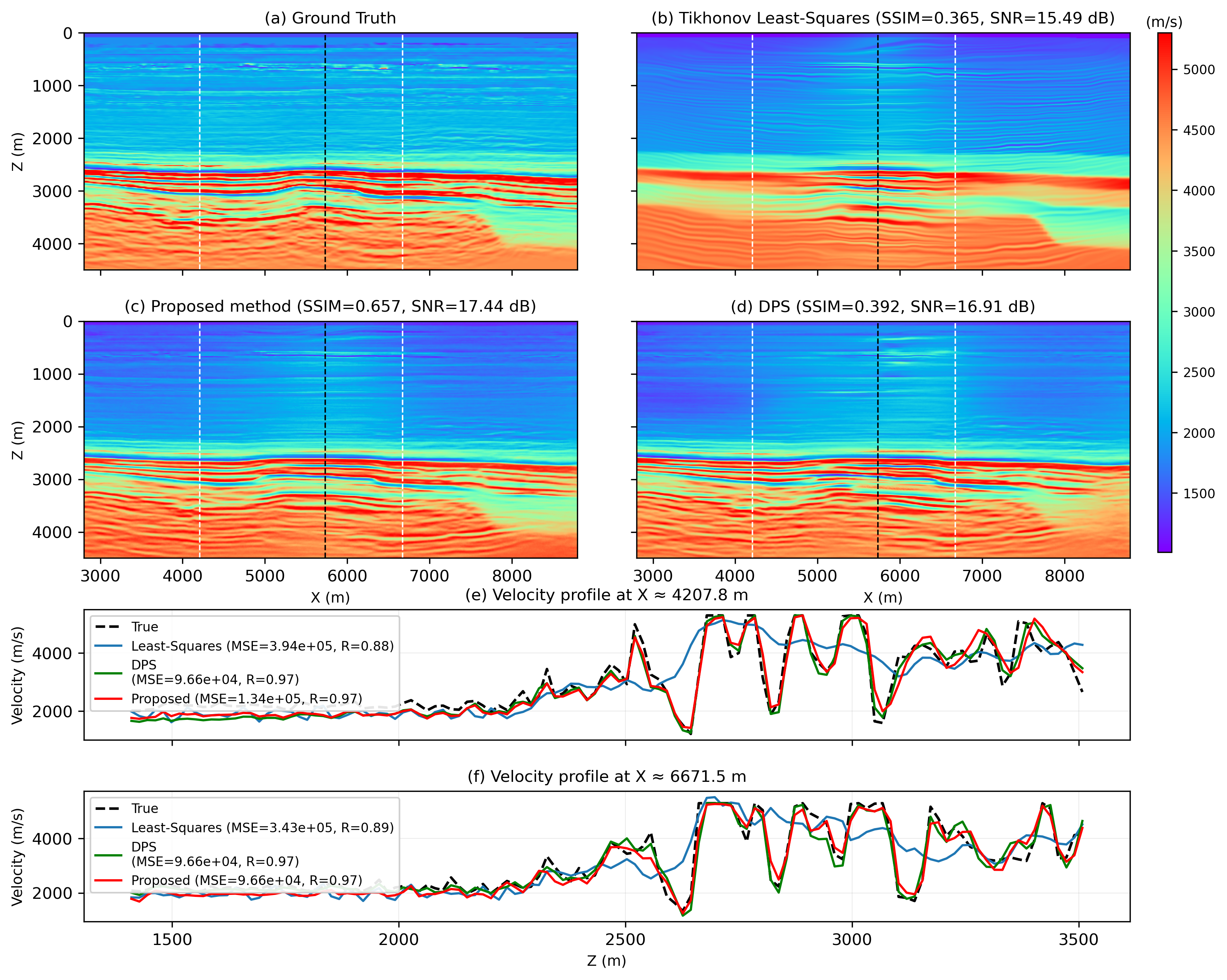}
    \caption{Velocity-model building (VMB) comparison for the synthetic test.
(a) Original (ground-truth) velocity model.
(b) Velocity model obtained from Tikhonov-regularized least-squares inversion.
(c) Velocity-model building result obtained with the proposed diffusion-based method.
(d) Velocity-model building result obtained using Diffusion Posterior Sampling (DPS).
The two white dashed lines indicate the locations at which velocity profiles are extracted, while the black dashed line marks the well position.
(e) Velocity profile extracted at $x = 4911.7 \mathrm{m}$.
(f) Velocity profile extracted at $x = 6671.5 \mathrm{m}$.}
    \label{fig:synth_results_vmb}
\end{figure}
To improve numerical stability, sampling is warm-started from the background model and initiated at $t/T = 0.1$, such that only the final 10\% of the reverse diffusion trajectory is executed. With $T=600$ total reverse steps, this corresponds to 60 denoising steps, of which 30 are guided. Since each guided step involves two LSQR iterations, the total number of LSQR iterations per edited sample is $N_{\mathrm{LSQR}} = 60$. The total runtime of this procedure is approximately 53 seconds.
For comparison, we also solve the inverse problem in Eq.~\ref{eq:ls} directly on the entire model $\mathbf{x}=\mathbf{S}\mathbf{t}$, using the same background velocity (Figure~\ref{fig:synth_intro}a) as the initial guess, the same regularization parameter $\lambda$, and 100 LSQR iterations, for a runtime of about 1.5 seconds on CPU. 
The DDIM-guided editing result (Figure~\ref{fig:synth_results}c) successfully recovers fine-scale structures within the target region and achieves the highest structural similarity (SSIM = 0.568), outperforming both the background model and the Tikhonov-regularized least-squares solution (SSIM = 0.414). The inferior performance of the Tikhonov solution is expected, as this approach relies solely on local regularization and does not incorporate any learned prior on realistic velocity structures, limiting its ability to reconstruct high-resolution features beyond those directly injected through the well and local slope. The DPS approach attains a comparable SSIM value (SSIM = 0.570) with a runtime of 30 seconds, indicating that both DPS and the proposed DDIM-guided method effectively exploit the learned diffusion prior to enhance structural realism. 
In terms of SNR, all the diffusion-based approaches significantly improve over the conventional structurally preconditioned inversion result. However, since the edited region is relatively localized, the structurally propagated well information already provides an accurate reconstruction even in the absence of an explicit diffusion prior. Nevertheless, the two approaches incorporating the diffusion prior still achieve slightly higher SNR values, indicating an additional improvement in structural consistency and local model refinement within the edited region. The Pearson correlation coefficient $R$ and the mean squared error (MSE) computed along velocity profiles away from the well locations (Figures~\ref{fig:synth_results}h and~\ref{fig:synth_results}i) further confirm that diffusion-guided methods consistently reconstruct sharper interfaces and high-contrast layers, while maintaining structural consistency with the surrounding background model. For visualization purposes, the least-squares solution is computed on the full model but displayed only within the editing region, with the surrounding area filled by the background velocity. The ground-truth model is shown using the same convention. In contrast, both diffusion-based approaches are able to directly generate the velocity model within the editing region.\\
We next consider the problem of full velocity model building, using the same pretrained diffusion model. In this case, the latent representation obtained through diffusion inversion is not masked, as the entire model is subject to updating, and the resulting latent space can be seen in Figure~\ref{fig:synth_ddiminversion}b. At each reverse diffusion timestep, the update $\Delta \mathbf{v}_t$ is obtained by solving the Tikhonov-image regularized system in Eq.~\ref{eq:obj_function_imreg} using two LSQR iterations, an image regularization weight $\kappa = 0.001$, and a damping parameter $\lambda = 0.01$. As in the editing case, $\Delta \mathbf{v}_t$ is saved and used as the initial guess for the LSQR solver at the subsequent timestep. Guidance is applied only during the last 50\% of the reverse diffusion process, the update is scaled by $\mu = 0.5$, and a reduced noise level $\eta = 0.1$ is used to introduce limited stochasticity during sampling.
To enhance stability, sampling is warm-started from the background model and initiated at $t/T = 0.3$, corresponding to the final 30\% of the DDIM inversion schedule, for a total of 43 seconds of runtime. For comparison, we also solve the full inverse problem in Eq.~\ref{eq:ls} directly on the entire model $\mathbf{v} = \mathbf{S}\mathbf{t}$, using the background model (Figure~\ref{fig:synth_results_vmb}d) as the initial guess, 50 LSQR iterations, image regularization weight $\kappa = 0.002$, and damping $\lambda = 0.01$ (in about 2 seconds of runtime).
The DDIM-based VMB result (Figure~\ref{fig:synth_results_vmb}c) leverages the learned prior to recover a velocity model that is closer to the ground truth, achieving a higher structural similarity (SSIM = 0.623) and signal-o-noise ratio (SNR=17.44 dB) than the imaging-regularized Tikhonov least-squares solution (SSIM = 0.365, SNR = 15.49 dB). The DPS approach closely resembles the result of the proposed method (SSIM=0.604, SNR = 61.91) and takes 33 seconds of GPU runtime. This result further highlights the importance of the learned diffusion prior: in both diffusion-based approaches, edge artifacts and migration crosstalk present in the RTM image and captured by the slope field are effectively suppressed, whereas these artifacts persist in the Tikhonov least-squares solution.  Consistent improvements are also observed in the Pearson coefficient $R$ and MSE computed along velocity profiles away from the well locations (Figures~\ref{fig:synth_results_vmb}h and~\ref{fig:synth_results_vmb}i), indicating that diffusion-guided sampling more effectively reconstructs high-contrast layers and sharper interfaces while preserving large-scale structural consistency.
\begin{figure}[h!]
    \centering
    \includegraphics[width=1.\linewidth]{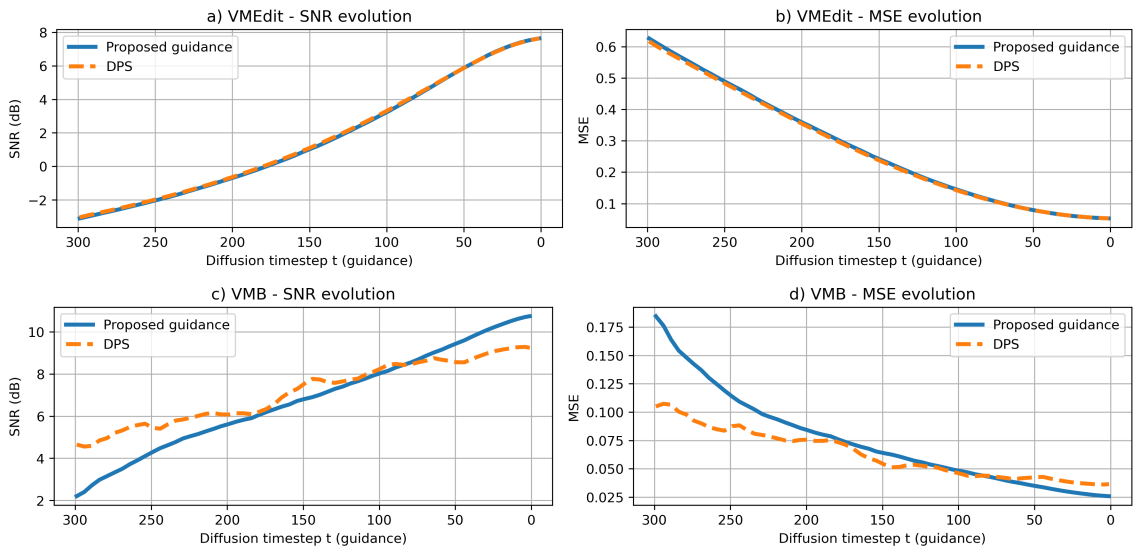}
    \caption{
Convergence analysis of the diffusion-based reconstruction approaches on the synthetic Volve experiments. 
(a) Evolution of the SNR during the VMEdit experiment. 
(b) Evolution of the MSE during the VMEdit experiment. 
(c) Evolution of the SNR during the VMB experiment. 
(d) Evolution of the MSE during the VMB experiment. 
The proposed guidance framework with iterative velocity-model updates is shown in blue, while the DPS curve is shown in orange.
}
\label{fig:metrics_evolution}
\end{figure}\\
The convergence curves in Figure \ref{fig:metrics_evolution} show that, in the VMEdit experiments, the MSE and SNR obtained by the two diffusion-guided approaches remain relatively close throughout the reverse diffusion process. This behavior is likely related to the localized nature of the editing task, where only a limited region of the velocity model is modified while the surrounding background model remains largely preserved. As a result, both guidance strategies are able to achieve comparable reconstruction quality. In contrast, larger differences in convergence behavior are observed in the VMB experiments, where the entire velocity model must be estimated. In this setting, the proposed guidance framework exhibits a more stable convergence trend and improved reconstruction accuracy compared to DPS. This behavior is attributed to the utilization of the LSQR-based guidance update, which provides a more stable and physically consistent estimation of the velocity model during the reverse diffusion process.

\subsection{Numerical examples on Viking Graben dataset}
\begin{figure}
    \centering
    \includegraphics[width=1.\linewidth]{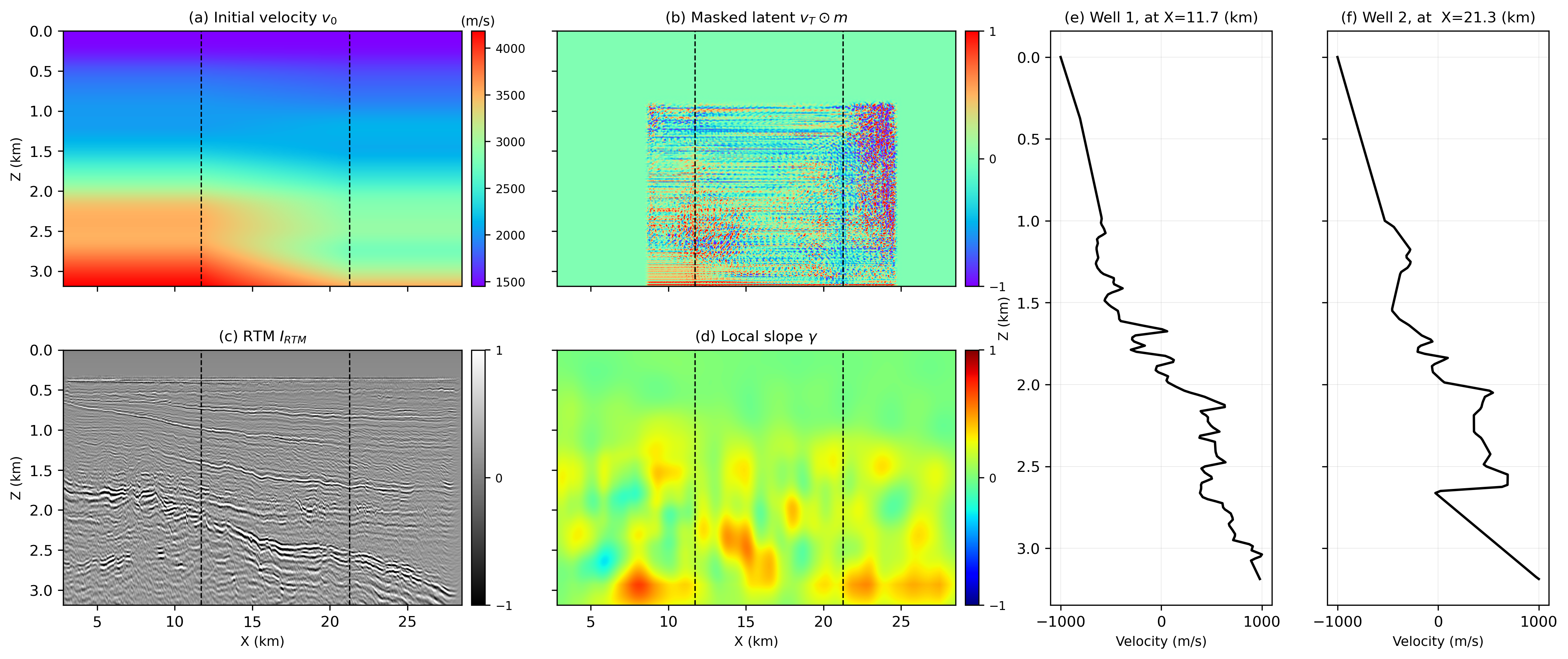}
    \caption{Viking Graben Dataset.
(a) Initial migration velocity model $\mathbf{v}_0$, where the two white dashed lines indicate the locations at which velocity profiles are extracted; the black dashed line marks the well position.
(b) Masked latent representation $\mathbf{v}_T$ obtained through DDIM inversion.
(c) RTM image used to derive structural guidance.
(d) Local slope field $\boldsymbol{\gamma}$ estimated from the RTM image.
(e) Well-log 1 positioned at $x = 11.7\mathrm{km}$, 
(f)  Well-log 1 positioned at $x = 21.3 \mathrm{km}$.}
    \label{fig:field_intro}
\end{figure}

We next demonstrate the applicability of the proposed diffusion-based framework on field data from the Viking Graben in the North Sea Basin. Specifically, we consider a 2D seismic line (Line 12) from the publicly available Mobil AVO Viking Graben dataset, which can be accessed through Mobil AVO Viking Graben 615 line 12 dataset (SEG Wiki). The dataset comprises 1001 shot gathers, each containing 6~seconds of recording time and 120 receivers, with temporal and common-midpoint sampling of 4~ms and 12.5~m, respectively \cite{Madiba2003}. 
This dataset presents several challenges for velocity-model building. In particular, the maximum source–receiver offset is relatively short (3237~m), and the water depth is shallow (approximately 350~m), resulting in seismic records that are strongly affected by water-bottom multiples and diffraction events. These factors limit illumination at depth and make the inversion problem particularly ill-posed. From the provided well-log measurements, we computed seismic-resolution ($V_p$) logs by applying Backus averaging and using linear interpolation to fill missing values. The resulting velocity profiles, referred to in this work as Well~1 and Well~2 (corresponding to Well-log 4 and Well-log 5 in the original dataset; Figures \ref{fig:field_intro}e and \ref{fig:field_intro}f), provide complementary constraints at different spatial locations.
In the following, we use this dataset to assess the robustness of the proposed diffusion-guided approach under realistic acquisition conditions and strong data limitations.
The initial velocity model $\mathbf{v}_0$ (Figure~\ref{fig:field_intro}a) is constructed by simple linear interpolation of the available well logs along the horizontal direction, providing a smooth background model that honors the well information but lacks lateral structural detail. The two black dashed lines mark the well positions.
The initial model $\mathbf{v}_0$ is projected onto the latent space through DDIM inversion for $T=600$ time samples (for a runtime of 13 seconds), and a spatial mask is applied according to Eq.~\ref{eq:mask_editing}, yielding the masked latent representation $\mathbf{v}_T$ shown in Figure~\ref{fig:field_intro}b. Structural guidance is derived from the RTM image (Figure \ref{fig:field_intro}c), computed using the initial velocity model. Local slopes $\pmb{\gamma}$, estimated from the RTM image using plane-wave destruction, are used to construct the structural operator that propagates well information into the surrounding regions (Figure~\ref{fig:field_intro}d).
Figures~\ref{fig:field_intro}e and~\ref{fig:field_intro}f show the two well logs used in this study, referred to as Well~1 and Well~2, positioned at $x = 11.7 \mathrm{km}$ and $x = 21.3 \mathrm{km}$, respectively. These well logs provide complementary velocity constraints at different spatial resolutions and are used to guide the diffusion-based velocity-model editing.
\begin{figure}[h!]
    \centering
    \includegraphics[width=1.\linewidth]{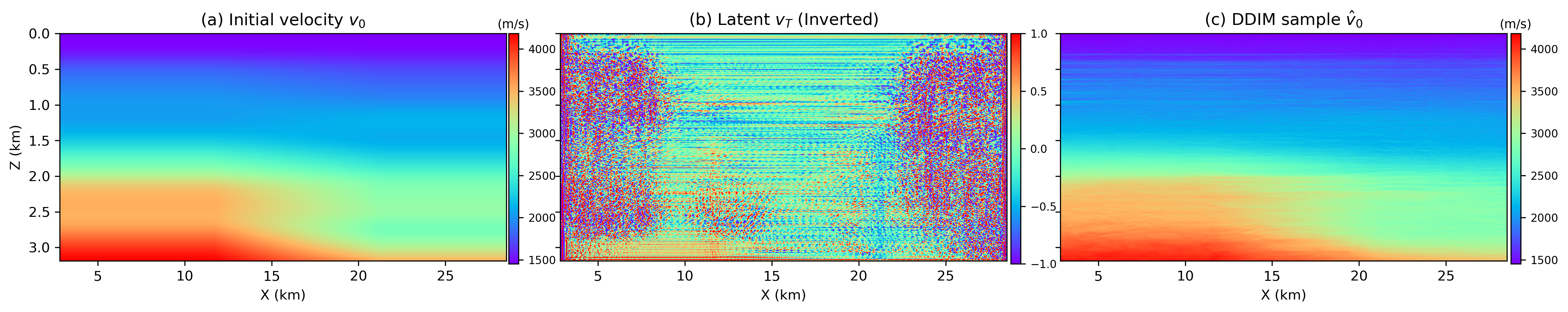}
    \caption{(a) Migration velocity model $\mathbf{v}_0$.
(b) Corresponding latent representation $\mathbf{v}_T$ obtained through DDIM inversion.
(c) Reconstructed velocity model $\hat{\mathbf{v}}_0$ recovered by deterministic DDIM sampling.}
    \label{fig:field_ddiminversion}
\end{figure}
Figure~\ref{fig:field_ddiminversion} summarizes the effect of DDIM inversion and deterministic sampling on the field-data background velocity model. \\
The interpolated initial model $\mathbf{v}_0$ (Figure~\ref{fig:field_ddiminversion}a) is projected into the latent space through DDIM inversion (Figure~\ref{fig:field_ddiminversion}b) and subsequently reconstructed via deterministic DDIM sampling to obtain $\hat{\mathbf{v}}_0$ (Figure~\ref{fig:field_ddiminversion}c), for 21 seconds of runtime for each procedures. Since no data-driven or imaging-based guidance is applied, the reconstructed model reflects the contribution of the learned diffusion prior alone, which enriches the background model with sharper interfaces and increased structural detail while remaining consistent with the available well-log information in which the velocity model was built using.
\begin{figure}
    \centering
    \includegraphics[width=1.\linewidth]{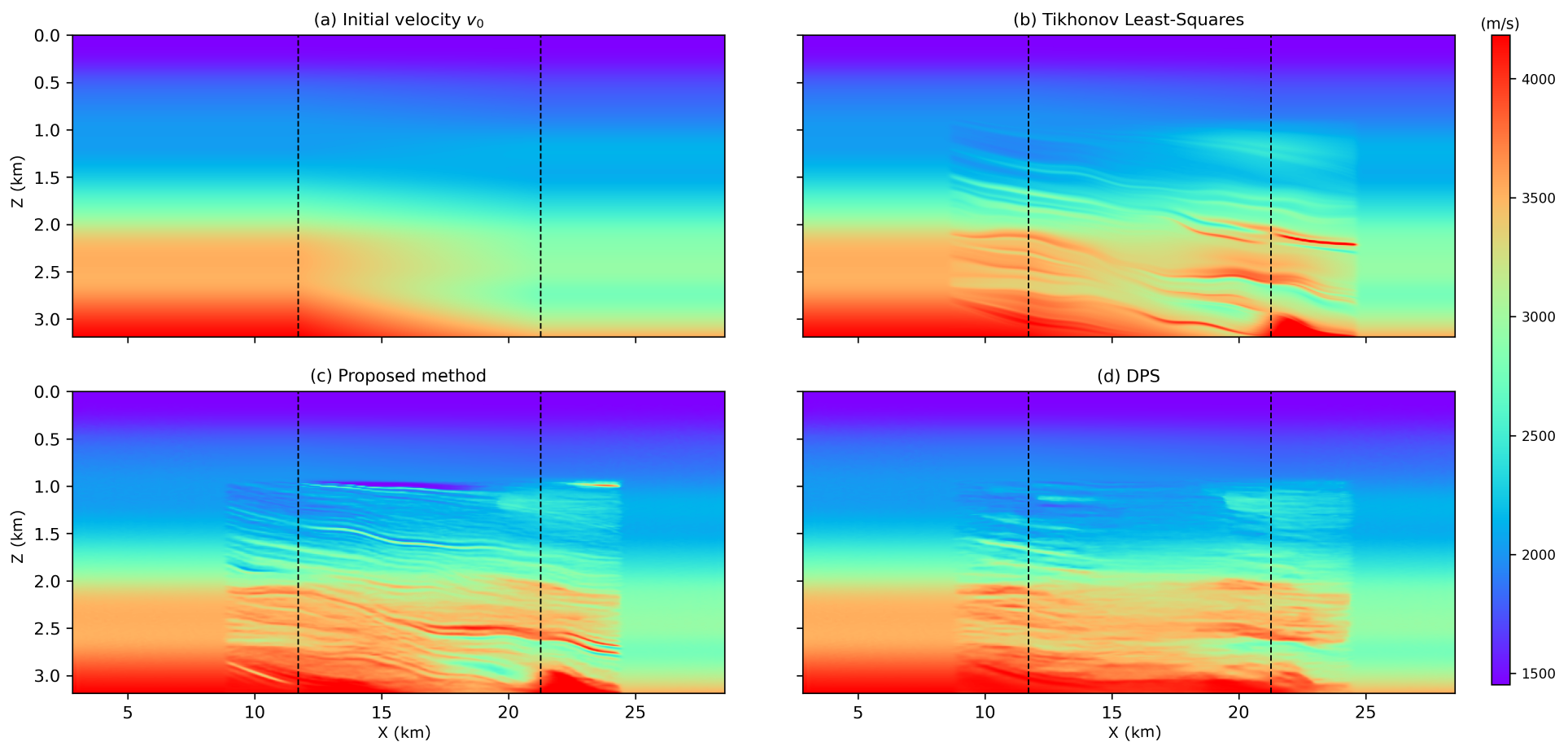}
    \caption{Velocity-model editing (VME) comparison for the Viking Graben Dataset.
(a) Initial migration velocity model.
(b) Conventional Tikhonov-regularized least-squares inversion.
(c) Velocity-model editing result obtained with the proposed diffusion-guided approach.
(d) Editing result obtained using the Diffusion Posterior Sampling (DPS) framework.
The two black dashed lines indicate the locations ($x = 11.7 \mathrm{km}$ and $x = 21.27 \mathrm{km}$) at which velocity profiles are extracted, while the black dashed line marks the well position}
    \label{fig:field_editing_results}
\end{figure}
Figure~\ref{fig:field_editing_results} compares velocity-model editing (VME) results for the Viking Graben dataset obtained with different inversion strategies. As in the synthetic experiments, the conventional Tikhonov-regularized least-squares solution (Figure~\ref{fig:field_editing_results}b) provides a smooth and geologically consistent interpolation of the well-log information, guided by the structural operator derived from the estimated slope field. This approach effectively propagates the well velocity information laterally while respecting the dominant structural trends inferred from the RTM image.
The proposed diffusion-guided editing method (Figure~\ref{fig:field_editing_results}c) closely follows the solution of the underlying inverse problem, and yields results that are largely consistent with the Tikhonov solution. However, the influence of the learned diffusion prior is clearly visible, leading to sharper velocity interfaces and enhanced fine-scale structure while preserving consistency with both the well data and the surrounding background model. This indicates that the diffusion prior acts as a complementary regularization mechanism that enriches the solution beyond what can be achieved through imaging-based regularization alone.
The Diffusion Posterior Sampling (DPS) result (Figure~\ref{fig:field_editing_results}d) also leverages the well-log constraints and the structural guidance provided by the slope field, while incorporating the learned high-resolution diffusion prior. In this field-data example, the DPS solution appears more strongly influenced by the prior, and less guided by the inverse problem solution. In terms of computational cost, the Tikhonov least-squares inversion requires approximately 1.5 seconds on CPU, while the proposed diffusion-guided approach and Diffusion Posterior Sampling (DPS) require about 51 seconds and 34 seconds on GPU, respectively.
\begin{figure}
    \centering
    \includegraphics[width=1.\linewidth]{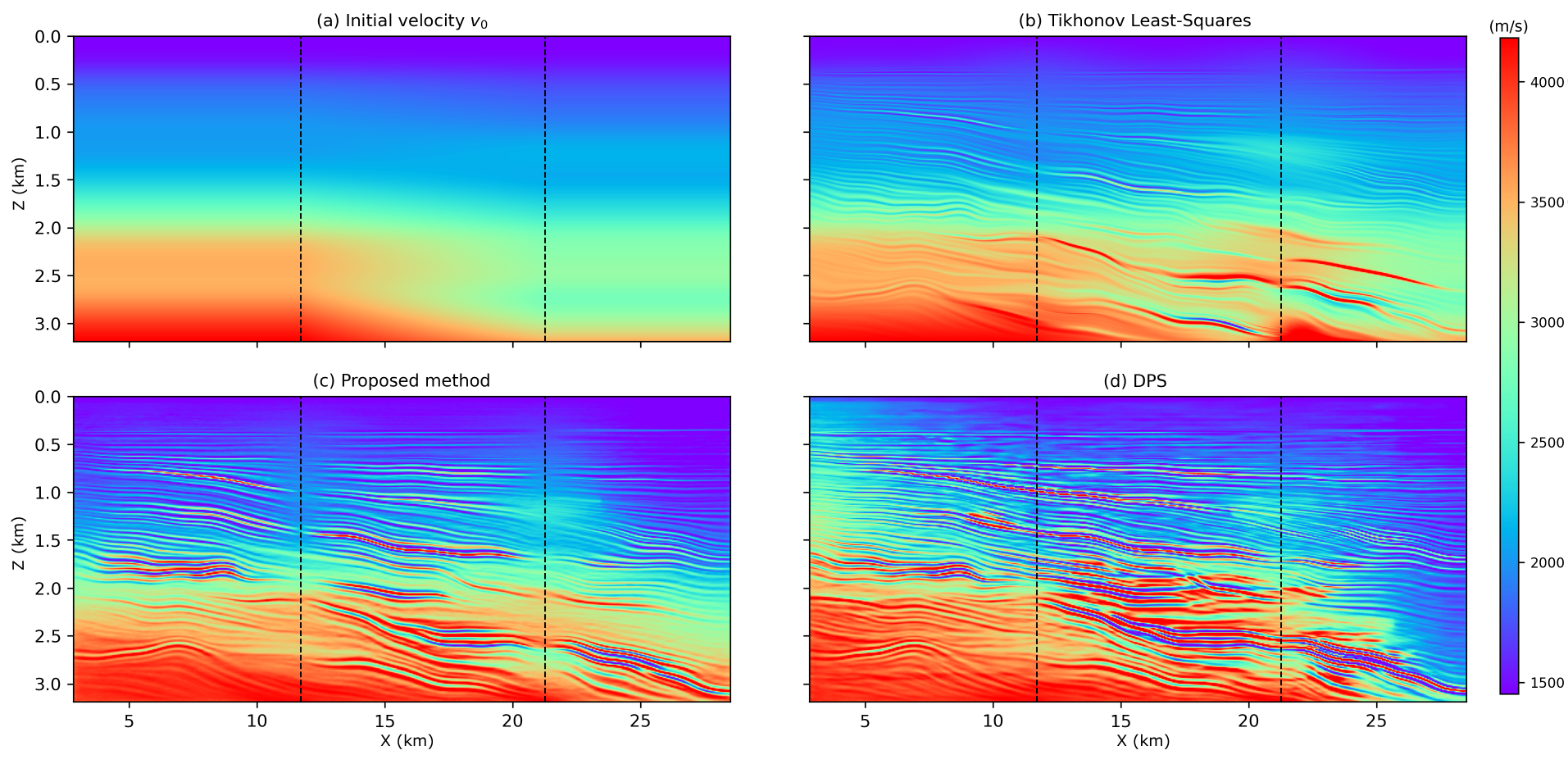}
    \caption{Velocity-model building (VMB) comparison for the Viking Graben Dataset.
(a) Original (ground-truth) velocity model.
(b) Velocity model obtained from Tikhonov-regularized least-squares inversion.
(c) Velocity-model building result obtained with the proposed diffusion-based method.
(d) Velocity-model building result obtained using Diffusion Posterior Sampling (DPS).
The two black dashed lines indicate the locations ($x = 11.7 \mathrm{km}$ and $x = 21.27 \mathrm{km}$) at which velocity profiles are extracted, while the black dashed line marks the well position.
}
    \label{fig:field_vmb_results}
\end{figure}
Figure~\ref{fig:field_vmb_results} compares full velocity model building (VMB) results for the Viking Graben dataset obtained using different inversion strategies. The Tikhonov-regularized least-squares solution (Figure~\ref{fig:field_vmb_results}b), similarly to the editing case, reconstructs the main sharp velocity interfaces in the vicinity of the wells. The inclusion of imaging-based regularization plays a key role in propagating structural information away from the immediate zone of influence of the structural smoothing operator, allowing coherent features to extend laterally into less constrained regions.
The proposed diffusion-based VMB result (Figure~\ref{fig:field_vmb_results}c) further improves the reconstruction by producing cleaner and sharper velocity interfaces. While imaging regularization continues to guide the large-scale structure, the learned diffusion prior enriches the solution with additional fine-scale details, leading to a more resolved and geologically consistent velocity model across the entire section.
The Diffusion Posterior Sampling (DPS) result (Figure~\ref{fig:field_vmb_results}d) also successfully recovers the overall velocity structure and benefits from both the well-log constraints and the learned diffusion prior. However, localized high-velocity artifacts are observed near the second well location, indicating a stronger influence of the prior in this region compared to the proposed DDIM-guided approach. Runtimes are approximately 1.5 seconds on CPU for least-squares inversion and 45 and 34 seconds on GPU for the proposed method and DPS, respectively.\\
\begin{figure}
    \centering
    \includegraphics[width=1.\linewidth]{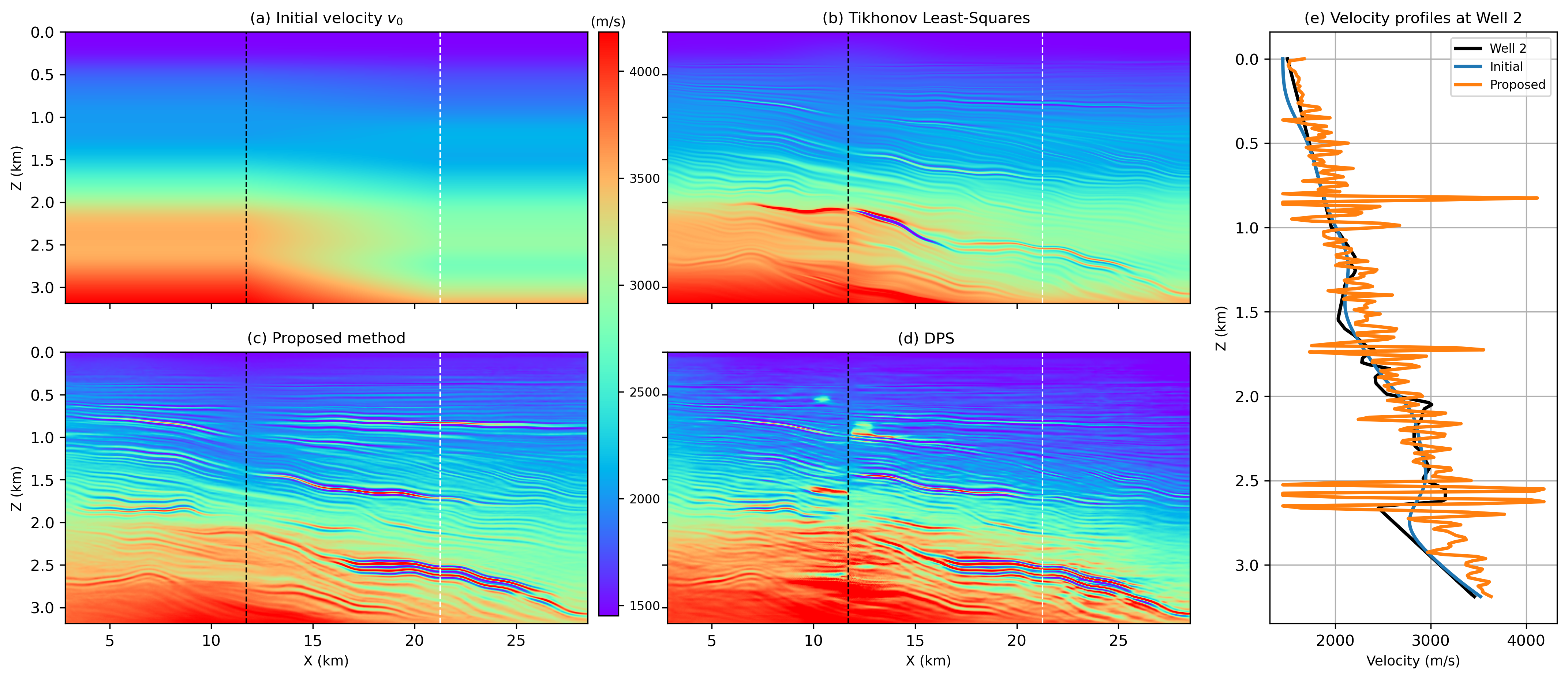}
    \caption{Blind-well validation for the Viking Graben field-data example. Well 2 (white dashed line) was excluded from the inversion and used only for validation. (a) Initial velocity model. (b) Least-squares reconstruction. (c) Proposed diffusion-guided reconstruction. (d) DPS reconstruction. (e) Velocity profiles extracted at the withheld well location, showing improved agreement with the unseen well log compared to the initial model.}
    \label{fig:field_vmb_results_well2}
\end{figure}
To provide a more stringent validation of the field-data experiment, we performed a blind-well test in which Well 2 was excluded from the inversion and used solely for validation purposes. The velocity reconstruction was therefore constrained only by the remaining well information, the imaging-based regularization, and the diffusion prior.\\
Figure \ref{fig:field_vmb_results_well2} shows the reconstructed velocity models together with the velocity profile extracted at the location of the excluded well. While the original evaluation focused on geological plausibility and image quality, this experiment provides an independent assessment of the reconstructed velocity model. As shown in Figure \ref{fig:field_vmb_results_well2}(e), the proposed diffusion-guided method follow the overall trend of the withheld well log and improve upon the initial velocity model. The proposed method additionally recovers higher-resolution structural variations that are consistent with the surrounding geological features inferred from the seismic image. For clarity, only the proposed diffusion-guided reconstructions is compared against the well log. The corresponding DPS reconstruction exhibits a very similar behavior at the validation well, and including all profiles in the same panel would unnecessarily reduce the readability of the figure . Overall, the blind-well experiment indicates that the reconstructed velocity models are not only geologically plausible but also remain consistent with independent well information that was not used during the inversion process.\\

\subsection{Ablation study}
To further investigate the role of structural guidance, we perform an ablation study on the Volve synthetic dataset in which the slope field used for guidance is computed directly from the ground-truth velocity model rather than from the RTM image. 
\begin{figure}
    \centering
    \includegraphics[width=1.\linewidth]{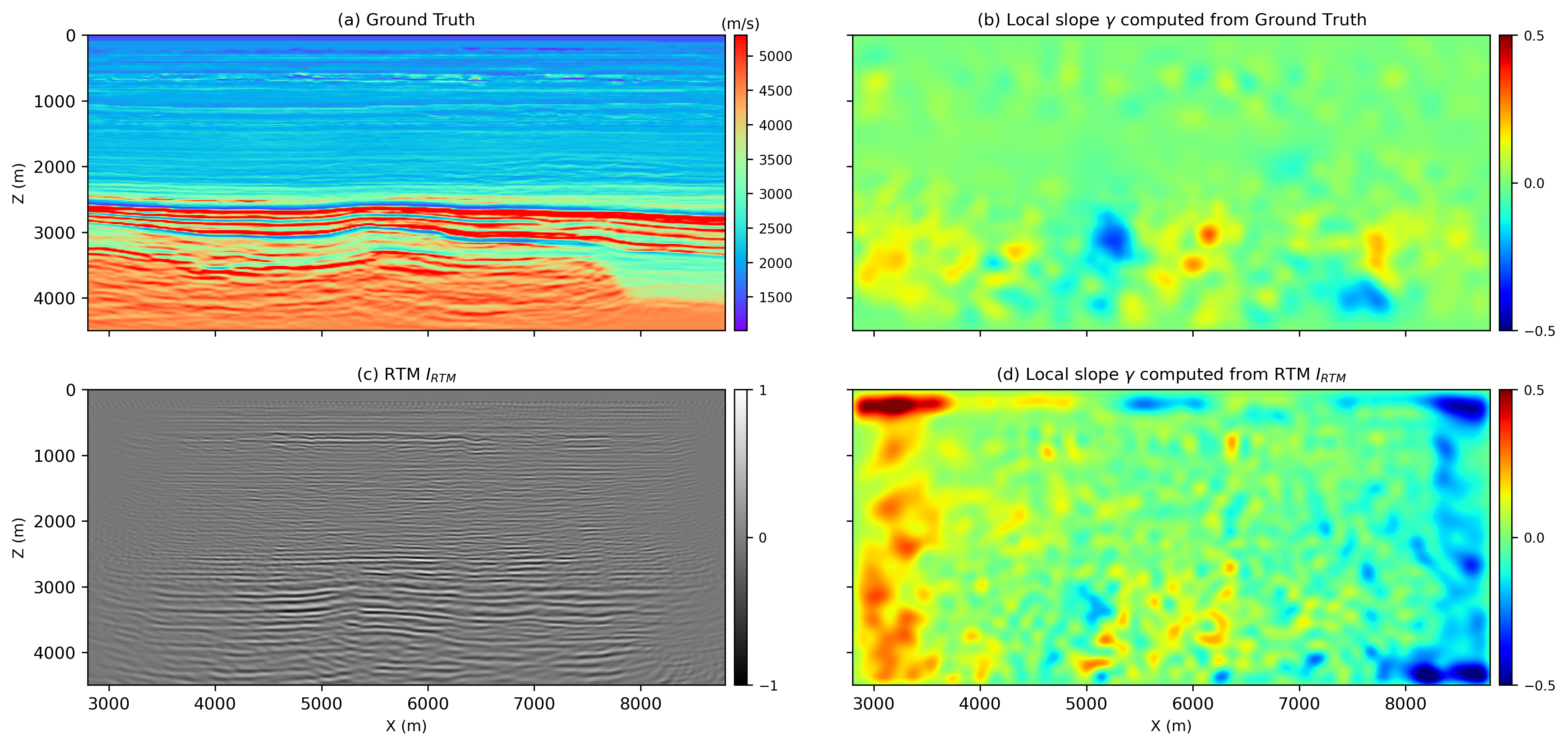}
    \caption{
(a) Ground-truth velocity model.
(b) Local slope field $\boldsymbol{\gamma}$ computed from the ground truth.
(c) Reverse Time Migration image $I_{\mathrm{RTM}}$.
(d) Local slope field $\boldsymbol{\gamma}$ computed from $I_{\mathrm{RTM}}$.}
    \label{fig:ablation_intro}
\end{figure}
Figure~\ref{fig:ablation_intro} compares the ground-truth velocity model (Figure~\ref{fig:ablation_intro}a) with the corresponding local slope field $\boldsymbol{\gamma}$ computed directly from the true model (Figure~\ref{fig:ablation_intro}b), as well as the RTM image (Figure~\ref{fig:ablation_intro}c) and the slope field estimated from it (Figure~\ref{fig:ablation_intro}d).
By replacing the RTM-derived slope with a more accurate slope field obtained from the ground-truth velocity, this experiment isolates the impact of slope-estimation accuracy on the inversion process. The ablation study allows us to assess how improved structural guidance influences both the Tikhonov least-squares solution and the diffusion-guided approaches, and to quantify the extent to which limitations in RTM-based slope estimation affect the final velocity reconstruction.
\begin{figure}
    \centering
    \includegraphics[width=1.\linewidth]{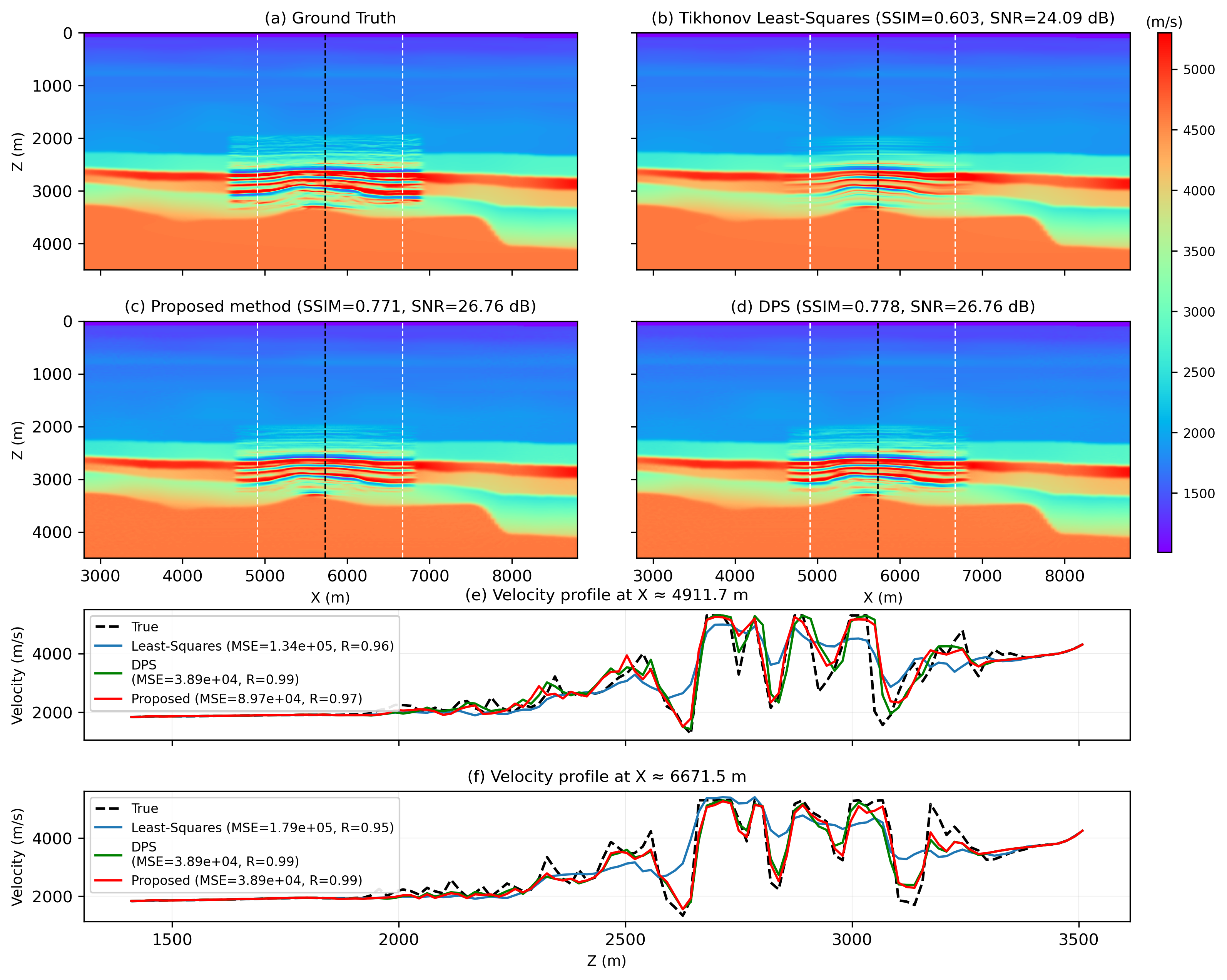}
    \caption{Velocity-model editing (VME) comparison for the synthetic test with slopes computed from the Ground Truth velocity model.
(a) Original (ground-truth) velocity model.
(b) Velocity model obtained from Tikhonov-regularized least-squares inversion.
(c) Velocity-model building result obtained with the proposed diffusion-based method.
(d) Velocity-model building result obtained using Diffusion Posterior Sampling (DPS).
The two white dashed lines indicate the locations at which velocity profiles are extracted, while the black dashed line marks the well position.
(e) Velocity profile extracted at $x = 4911.7,\mathrm{m}$.
(f) Velocity profile extracted at $x = 6671.5,\mathrm{m}$.
In panels (e) and (f), the black dashed curve denotes the ground-truth velocity profile, the blue curve corresponds to the Tikhonov least-squares result, the green curve to the DPS result, and the red curve to the proposed method.}
    \label{fig:synth_edit_results_ablation}
\end{figure}
Figure~\ref{fig:synth_edit_results_ablation} presents the velocity-model editing (VME) results for the ablation study on the Volve synthetic dataset, where the structural guidance is derived from slopes computed directly from the ground-truth velocity model. In this idealized setting, all approaches achieve a high level of structural similarity with the reference model, confirming that accurate slope information provides an effective preconditioning of the inverse problem.
The Tikhonov-regularized least-squares solution (Figure~\ref{fig:synth_edit_results_ablation}b) benefits significantly from the improved structural guidance, allowing sharp interfaces and coherent layering to be recovered well beyond the immediate vicinity of the well location. Similarly, both diffusion-based approaches, the proposed DDIM-guided method (Figure~\ref{fig:synth_edit_results_ablation}c) and Diffusion Posterior Sampling (DPS) (Figure~\ref{fig:synth_edit_results_ablation}d) exploit the high-quality slope field to guide the reconstruction, resulting in velocity models that closely match the ground truth.
The velocity profiles extracted at $x = 4911.7,\mathrm{m}$ and $x = 6671.5,\mathrm{m}$ (Figures~\ref{fig:synth_edit_results_ablation}e and~\ref{fig:synth_edit_results_ablation}f) further illustrate this behavior. All methods accurately recover the main velocity contrasts and layer boundaries, with the improved slope information facilitating both the least-squares inversion and the diffusion-guided sampling. These results highlight that the availability of accurate structural slopes strongly enhances inversion performance and reduces the relative differences between conventional and diffusion-based approaches, while still enabling the diffusion prior to contribute to the recovery of fine-scale details. Figure~\ref{fig:synth_edit_results_ablation} presents the velocity-model building results for the ablation study on the Volve synthetic dataset, where the structural guidance is derived from slopes computed directly from the ground-truth velocity model. In this idealized setting, all approaches achieve a high level of structural similarity and SNR with the reference model, confirming that accurate slope information provides an effective preconditioning of the inverse problem.
The Tikhonov-regularized least-squares solution (Figure~\ref{fig:synth_edit_results_ablation}b) benefits significantly from the improved structural guidance, allowing sharp interfaces and coherent layering to be recovered well beyond the immediate vicinity of the well location. Similarly, both diffusion-based approaches, the proposed DDIM-guided method (Figure~\ref{fig:synth_edit_results_ablation}c) and Diffusion Posterior Sampling (DPS) (Figure~\ref{fig:synth_edit_results_ablation}d), exploit the high-quality slope field to guide the reconstruction, resulting in velocity models that closely match the ground truth.
Importantly, the use of ground-truth-derived slopes enables the recovery of subtle structural features, such as the small fold located in the central part of the editing region, which was not clearly extracted when slopes were computed from the RTM image due to imaging noise and crosstalks masking this structure. The velocity profiles extracted at $x = 4911.7,\mathrm{m}$ and $x = 6671.5,\mathrm{m}$ (Figure~\ref{fig:synth_edit_results_ablation}e and~\ref{fig:synth_edit_results_ablation}f) further confirm that all methods accurately reconstruct the main velocity contrasts and layer boundaries. These results demonstrate that improved slope accuracy not only enhances inversion performance across all approaches but also strengthens the effectiveness of diffusion-guided sampling by providing more reliable structural guidance.
\begin{figure}
    \centering
    \includegraphics[width=1.\linewidth]{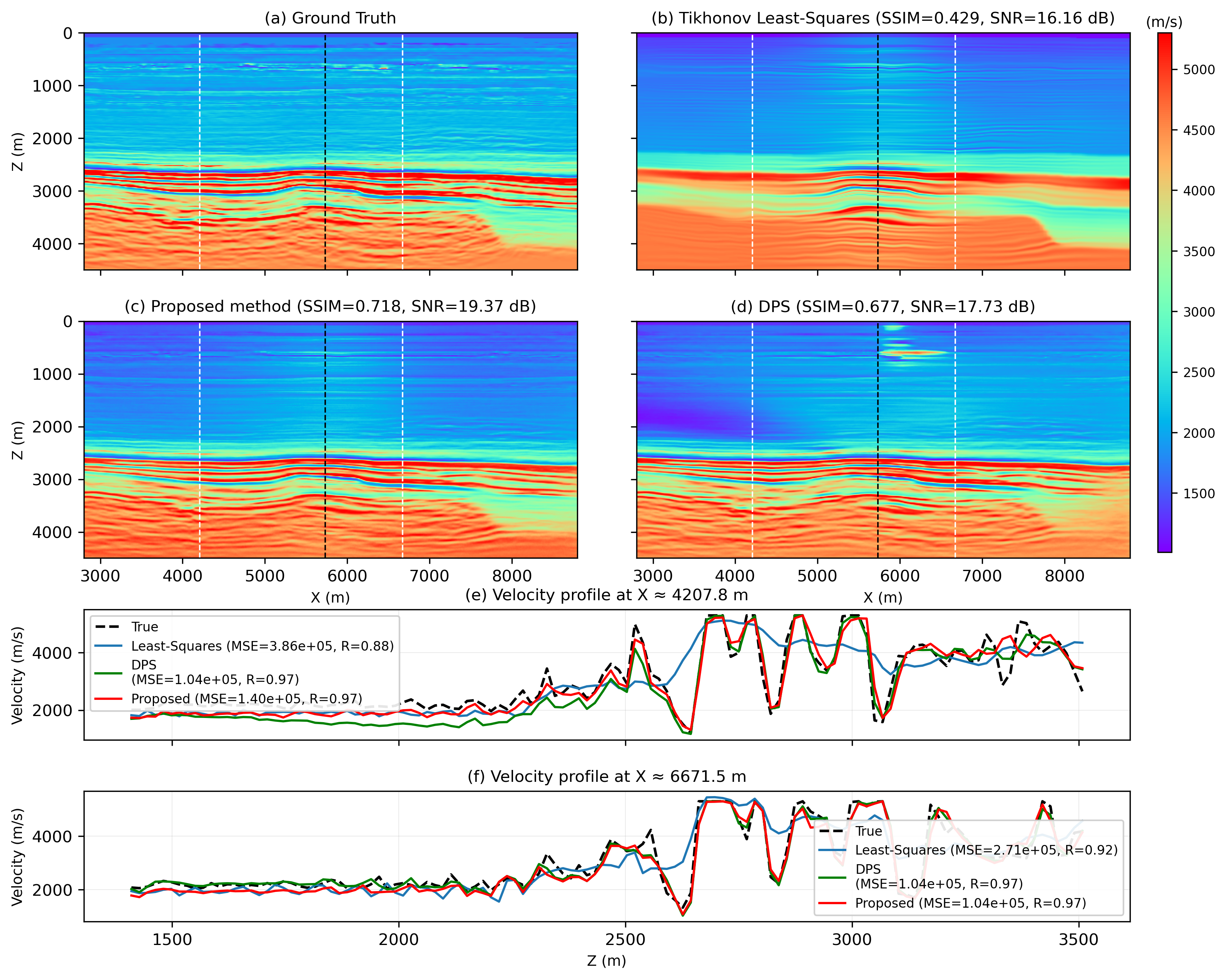}
    \caption{Velocity-model building (VMB) comparison for the synthetic test.
(a) Original (ground-truth) velocity model.
(b) Velocity model obtained from Tikhonov-regularized least-squares inversion.
(c) Velocity-model building result obtained with the proposed diffusion-based method.
(d) Velocity-model building result obtained using Diffusion Posterior Sampling (DPS).
The two white dashed lines indicate the locations at which velocity profiles are extracted, while the black dashed line marks the well position.
(e) Velocity profile extracted at $x = 4911.7,\mathrm{m}$.
(f) Velocity profile extracted at $x = 6671.5,\mathrm{m}$.
In panels (e) and (f), the black dashed curve denotes the ground-truth velocity profile, the blue curve corresponds to the Tikhonov least-squares result, the green curve to the DPS result, and the red curve to the proposed method.}
    \label{fig:synth_vmb_results_ablation}
\end{figure}
Figure~\ref{fig:synth_vmb_results_ablation} shows the results of the velocity model building (VMB) ablation study on the Volve synthetic dataset, where structural guidance is derived from slopes computed directly from the ground-truth velocity model. Consistent with the localized editing case, all approaches benefit from the more accurate slope information, leading to improved structural accuracy in the vicinity of the well and enhanced propagation of structural features away from the well location.
Both diffusion-based methods—the proposed DDIM-guided approach (Figure~\ref{fig:synth_vmb_results_ablation}c) and Diffusion Posterior Sampling (DPS) (Figure~\ref{fig:synth_vmb_results_ablation}d), recover coherent velocity structures and clearly resolve the small fold in the central part of the model, which is only partially visible when slopes are estimated from the RTM image. The Tikhonov-regularized least-squares solution (Figure~\ref{fig:synth_vmb_results_ablation}b) also reconstructs the main structural features and yields a particularly smooth and clean velocity model.
The velocity profiles extracted at $x = 4911.7,\mathrm{m}$ and $x = 6671.5,\mathrm{m}$ (Figures~\ref{fig:synth_vmb_results_ablation}e and~\ref{fig:synth_vmb_results_ablation}f) confirm that all methods accurately recover the dominant velocity contrasts and layer boundaries. Overall, these results indicate that improved slope accuracy enhances VMB performance across all approaches, while also highlighting the different trade-offs between structural fidelity and smoothing inherent to diffusion-based and least-squares methods. As a result, we will investigate in the future using diffusion models to generate guided slopes instead of relying on the PWD approach.\\

\section{Discussion}
The results consistently show that the diffusion-based approaches outperform the conventional Tikhonov least-squares (LS) inversion framework in both velocity-model editing (VME) and velocity model building (VMB) tasks. The learned diffusion prior plays a central role in enhancing the solution of the inverse problem, enabling the recovery of sharper interfaces and more realistic high-resolution structures than those obtained through classical regularization alone.\\
An important aspect of diffusion-based inversion is the construction of the training dataset used to learn the prior. Since the diffusion prior strongly influences the sampling process, its quality and representativeness must be carefully evaluated. A well-designed prior leads to improved reconstruction quality, whereas a poorly chosen or mismatched prior may result in solutions that are inferior to those obtained with conventional least-squares approaches. In this sense, the selection and design of the training dataset is of fundamental importance. In industrial settings, this limitation can be mitigated by tailoring the training data to the specific geological context of the area of interest, leveraging prior geological knowledge and available well-log information to build highly consistent, high-resolution velocity-model datasets. In contrast, in this work, we rely on publicly available open-source velocity models, which may limit the expressiveness of the learned prior.\\
When comparing the two diffusion-based approaches, the proposed DDIM-guided method generally achieves slightly better or comparable results relative to Diffusion Posterior Sampling (DPS), although the overall performance of the two methods remains very close across all experiments. A notable advantage of the proposed approach is its relative simplicity in terms of hyperparameter tuning and control. DPS appears to rely more strongly on the diffusion prior, likely due to the explicit use of the score-network gradient in the sampling update, which injects the prior more aggressively into the solution. In contrast, the proposed method places greater emphasis on the solution of the underlying inverse problem, allowing the data-driven guidance to play a more balanced role in the final reconstruction.\\
From a computational perspective, DPS exhibits a slightly faster sampling runtime, requiring approximately 33 seconds in our experiments, compared to 45–53 seconds for the proposed DDIM-guided approach. This difference arises because the proposed method explicitly solves a Tikhonov-regularized inverse problem at each guided diffusion step using LSQR, whereas DPS relies solely on gradient-based updates during sampling, which are computationally more efficient.\\
The guidance for both approaches was intentionally applied only during the final portion of the reverse diffusion trajectory. Early diffusion steps correspond to highly noisy latent states, where strong physical guidance may destabilize the generative trajectory and interfere with the large-scale structural prior learned by the diffusion model. By contrast, late-stage guidance acts on partially denoised samples that already contain coherent geological structures, allowing the guidance term to refine the reconstruction while preserving the realism and continuity imposed by the learned prior. In practice, the selected guidance stages were chosen empirically as a compromise between reconstruction stability, computational cost, and effective enforcement of the inversion constraints. Applying the guidance too early or over the entire diffusion trajectory was observed to increase instability and reduce the generative flexibility of the model. Another important parameter to take inot consideration is $\mu$, which controls the strength of the guidance contribution and therefore governs the balance between the learned diffusion prior and the inversion-driven constraints. Smaller values of $\mu$ lead to weaker guidance and produce solutions that remain closer to the unconditional generative prior, whereas larger values increase the influence of the physical and structural constraints. However, excessively large $\mu$ values may over-constrain the reconstruction, destabilize the reverse process, and reduce geological realism. In the experiments presented in this work, $\mu = 0.5$ was selected empirically as a stable compromise between reconstruction fidelity, structural consistency, and convergence stability.\\
The least-squares inversion framework remains competitive, particularly for localized editing tasks, where it can recover geologically plausible results by propagating well-log information through imaging-based regularization. For full velocity model building, the LS approach performs well in extending structural information away from the wells, producing smooth and clean solutions in which well-log constraints and imaging information are consistently merged into the background model. However, the absence of a learned prior limits its ability to reconstruct fine-scale details and suppress imaging artifacts.\\
Specifically, the imaging regularizer is not intended to reproduce the RTM amplitudes directly in the velocity model. Instead, it is used as a structural surrogate that encourages the reconstructed velocity model to contain vertical contrasts whose geometry is consistent with the reflector information present in the RTM image. In this sense, the operator provides a reflectivity-like constraint derived from the vertical derivative of the slowness-squared perturbation, and it is mainly used to guide the location and continuity of velocity interfaces rather than to impose a fully amplitude-consistent seismic imaging condition. In particular, strong migration artifacts, poor illumination, inaccurate background velocities, or phase/amplitude mismatch may reduce the reliability of the imaging constraint. For this reason, the imaging term is used as a regularization component rather than as a dominant data-fitting term, and its influence is balanced with the well constraints, structural preconditioning, and diffusion prior.\\
Finally, the ablation study highlights the critical importance of accurate structural slope information for both diffusion-based and least-squares inversion. When slopes are computed directly from the ground-truth velocity model, all methods achieve significantly improved structural accuracy, confirming that slope-based guidance acts as an effective preconditioner of the inverse problem. In practice, however, slope estimation quality is constrained by the fidelity of the RTM image and, ultimately, by the quality of the acquired seismic data, as well as the robustness of the approach used to estimate the slope. As a potential future direction, diffusion models could be trained not only to learn velocity-model priors but also to predict structural slopes from high-resolution velocity models, potentially improving guidance quality and further enhancing inversion robustness.

\section{Conclusions}
We have introduced a unified framework for velocity-model editing (VME) and velocity model building (VMB) that combines a learned diffusion prior with guided well information. For velocity-model editing, the proposed DDIM-guided approach integrates a diffusion prior learned from high-resolution velocity models with a well-based structurally preconditioned Tikhonov inversion, enabling the recovery of fine-scale features within the edited region while preserving the global character of the background velocity model. The editing results are systematically compared with both a conventional least-squares solution and Diffusion Posterior Sampling (DPS), showing that diffusion-based approaches provide superior structural realism. These comparisons are validated not only on synthetic benchmarks but also through field-data editing experiments on the Viking Graben dataset, demonstrating the practical relevance of the proposed method.
We then extend the framework to the more challenging problem of full velocity model building (VMB), formulating a structurally preconditioned well-matching inverse problem augmented with imaging-based regularization. This represents a contribution not only within the diffusion framework, but also for the conventional least-squares formulation, where imaging-based regularization significantly improves the lateral propagation of structural information away from the wells. The resulting inverse problem is solved using a classical Tikhonov least-squares approach, the proposed DDIM-guided method, and DPS. While all approaches benefit from the structural smoother in propagating well information, the diffusion-based solutions consistently yield superior results by leveraging the learned prior to recover structurally consistent, high-resolution features away from the wells. Comparisons between DDIM-guided sampling and DPS indicate comparable reconstruction quality, with the proposed approach offering improved controllability and robustness, while DPS exhibits slightly lower computational cost.
The robustness and generalization capability of the proposed framework are demonstrated through extensive numerical experiments on the Viking Graben dataset, for both velocity-model editing and full velocity model building, confirming that diffusion-based priors trained on synthetic data can be effectively transferred to field-scale applications. An ablation study further highlights that structural slope information plays a fundamental role in the final reconstruction quality, acting as a powerful preconditioner for both least-squares and diffusion-based methods. Improved slope accuracy leads to significant gains in structural fidelity and enables the recovery of subtle geological features that are otherwise obscured by imaging noise. Future work will therefore focus on improving slope estimation, potentially through learning-based approaches, and on further investigating posterior sampling strategies within the diffusion framework.

\section*{Acknowledgments}
This publication is based on work supported by the King Abdullah University of Science and Technology (KAUST). The authors thank the DeepWave sponsors for supporting this research.




\begin{thebibliography}{}

\bibitem[Alfarhan et~al., 2024]{Alfarhan2024}
Alfarhan, M., Ravasi, M., Chen, F., and Alkhalifah, T. (2024).
\newblock Robust full waveform inversion with deep hessian deblurring.
\newblock {\em Geophysical Journal International}, 240(1):303--316.

\bibitem[AlYahya, 1988]{AlYahya1988}
AlYahya, K.~M. (1988).
\newblock Velocity analysis by iterative profile migration.
\newblock SEG Technical Program Expanded Abstracts 1988:901--903.

\bibitem[Araya-Polo et~al., 2018]{Araya-Polo2018}
Araya-Polo, M., Jennings, J., Adler, A., and Dahlke, T. (2018).
\newblock Deep-learning tomography.
\newblock {\em The Leading Edge}, 37(1):58--66.

\bibitem[Ayeni et~al., 2009]{Ayeni2009}
Ayeni, G., Tang, Y., and Biondi, B. (2009).
\newblock Joint preconditioned least-squares inversion of simultaneous source
  time-lapse seismic data sets.
\newblock SEG Technical Program Expanded Abstracts 2009:3914--3918.

\bibitem[Bakulin et~al., 2010]{Bakulin2010}
Bakulin, A., Woodward, M., Nichols, D., Osypov, K., and Zdraveva, O. (2010).
\newblock Localized anisotropic tomography with well information in vti media.
\newblock {\em Geophysics}, 75(5):D37--D45.

\bibitem[bin Waheed et~al., 2021]{waheed2021}
bin Waheed, U., Alkhalifah, T., Haghighat, E., Song, C., and Virieux, J.
  (2021).
\newblock Pinntomo: Seismic tomography using physics-informed neural networks.

\bibitem[Biondi, 2005]{Biondi2005}
Biondi, B. (2005).
\newblock Angle-domain common image gathers for anisotropic migration.
\newblock SEG Technical Program Expanded Abstracts 2005:1922--1925.

\bibitem[Biondi et~al., 2023]{Biondi2023}
Biondi, E., Barnier, G., Biondi, B., and Clapp, R.~G. (2023).
\newblock Target-oriented elastic full-waveform inversion through acoustic
  extended image-space redatuming.
\newblock {\em Geophysics}, 88(3):R269--R296.

\bibitem[Bishop et~al., 1985]{Bishop1985}
Bishop, T.~N., Bube, K.~P., Cutler, R.~T., Langan, R.~T., Love, P.~L., Resnick,
  J.~R., Shuey, R.~T., Spindler, D.~A., and Wyld, H.~W. (1985).
\newblock Tomographic determination of velocity and depth in laterally varying
  media.
\newblock {\em Geophysics}, 50(6):903--923.

\bibitem[Bleistein et~al., 2001]{Bleistein2001}
Bleistein, N., Cohen, J.~K., and Stockwell, J.~W. (2001).
\newblock {\em Mathematics of Multidimensional Seismic Inversion}.
\newblock Springer.

\bibitem[Brandolin et~al., 2024]{Brandolin2024}
Brandolin, F., Ravasi, M., and Alkhalifah, T. (2024).
\newblock Pinnslope: Seismic data interpolation and local slope estimation with
  physics informed neural networks.
\newblock {\em GEOPHYSICS}, 89(4):V331--V345.

\bibitem[Chen et~al., 2016]{Chen2016}
Chen, Y., Chen, H., Xiang, K., and Chen, X. (2016).
\newblock Geological structure guided well log interpolation for high-fidelity
  full waveform inversion.
\newblock {\em Geophysical Journal International}, 207(2):1313--1331.

\bibitem[Chiu and Stewart, 1987]{Chiu1987}
Chiu, S. K.~L. and Stewart, R.~R. (1987).
\newblock Tomographic determination of three-dimensional seismic velocity
  structure using well logs, vertical seismic profiles, and surface seismic
  data.
\newblock {\em Geophysics}, 52(8):1085--1098.

\bibitem[Chung et~al., 2023]{chung2023dps}
Chung, H., Kim, J., McCann, M.~T., Klasky, S.~M., and Ye, J.~C. (2023).
\newblock Diffusion posterior sampling for general noisy inverse problems.
\newblock In {\em ICLR}.

\bibitem[Clapp et~al., 2004]{Clapp2004}
Clapp, R.~G., Biondi, B., and Claerbout, J.~F. (2004).
\newblock Incorporating geologic information into reflection tomography.
\newblock {\em Geophysics}, 69(2):533--546.

\bibitem[Cui et~al., 2020]{Cui2020}
Cui, T., Rickett, J., Vasconcelos, I., and Veitch, B. (2020).
\newblock Target-oriented full-waveform inversion using marchenko redatumed
  wavefields.
\newblock {\em Geophysical Journal International}, 223(2):792--810.

\bibitem[Dix, 1955]{Dix1955}
Dix, C.~H. (1955).
\newblock Seismic velocities from surface measurements.
\newblock {\em Geophysics}, 20(1):68--86.

\bibitem[E. and Hale, 2015]{Naeini2015}
E., N. and Hale, D. (2015).
\newblock Image- and horizon-guided interpolation.
\newblock {\em Geophysics}, 80:V47–V56.

\bibitem[Fomel, 2002]{Fomel2002}
Fomel, S. (2002).
\newblock Applications of plane-wave destruction filters.
\newblock {\em Geophysics}, 67:1946–1960.

\bibitem[Gebre et~al., 2025]{Gebre2025}
Gebre, M.~G., Pinheiro, D.~N., Duarte, E.~F., Costa, C. A. N.~d., Araújo, R.
  C.~F., Silva, D.~d., and Araújo, J. M.~d. (2025).
\newblock Preconditioning full wave inversion using geological
  structure-oriented gradient filtering.
\newblock {\em IEEE Geoscience and Remote Sensing Letters}, 22:1--5.

\bibitem[Geng et~al., 2021]{Zhicheng2021}
Geng, Z., Zhao, Z., Shi, Y., Wu, X., Fomel, S., and Sen, M. (2021).
\newblock Deep learning for velocity model building with common-image gather
  volumes.
\newblock {\em Geophysical Journal International}, 228(2):1054--1070.

\bibitem[Goren and Treister, 2024]{Goren2024}
Goren, M.~M. and Treister, E. (2024).
\newblock Physics-guided full waveform inversion using encoder-solver
  convolutional neural networks.
\newblock {\em Inverse Problems}, 40(12):125003.

\bibitem[Guitton et~al., 2012]{Guitton2012}
Guitton, A., Ayeni, G., and Díaz, E. (2012).
\newblock Constrained full-waveform inversion by model reparameterization.
\newblock {\em Geophysics}, 77(2):R117--R127.

\bibitem[Guo and Alkhalifah, 2020]{Guo2020}
Guo, Q. and Alkhalifah, T. (2020).
\newblock Target-oriented waveform redatuming and high-resolution inversion:
  Role of the overburden.
\newblock {\em Geophysics}, 85(6):R525--R536.

\bibitem[Hale, 2010]{Hale2010Imageguided3I}
Hale, D. (2010).
\newblock Image-guided 3d interpolation of borehole data.
\newblock {\em Seg Technical Program Expanded Abstracts}.

\bibitem[He and Wang, 2020]{He2020}
He, Q. and Wang, Y. (2020).
\newblock Reparameterized full-waveform inversion using deep neural networks.
\newblock {\em Geophysics}, 86(1):V1--V13.

\bibitem[Ho et~al., 2020]{Ho2020ddpm}
Ho, J., Jain, A., and Abbeel, P. (2020).
\newblock Denoising diffusion probabilistic models.
\newblock In {\em NeurIPS}, page (33).

\bibitem[Huang et~al., 2023]{Huang2023}
Huang, G., Crawley, S., Djebbi, R., Ramos-Martinez, J., and Chemingui, N.
  (2023).
\newblock Automated velocity model building using fourier neural operators.
\newblock SEG/AAPG International Meeting for Applied Geoscience \&
  Energy:SEG--2023--3911535.

\bibitem[Karimi et~al., 2017]{Karimi2017}
Karimi, P., Fomel, S., and Zhang, R. (2017).
\newblock Creating detailed subsurface models using predictive image-guided
  well-log interpolation.
\newblock {\em Interpretation}, 5(3):T279--T285.

\bibitem[Kazei et~al., 2020]{Kazei2020}
Kazei, V., Ovcharenko, O., and Alkhalifah, T. (2020).
\newblock Velocity model building by deep learning: From general synthetics to
  field data application.
\newblock SEG International Exposition and Annual Meeting:D041S100R002.

\bibitem[Krige, 1951]{Krige1951}
Krige, D.~G. (1951).
\newblock A statistical approach to some mine val- uation and allied problems
  on the witwatersrand: M.sc. dissertation.
\newblock {\em M.Sc. dissertation, University of the Witwatersrand.}

\bibitem[Li and Alkhalifah, 2022]{Li2022}
Li, Y. and Alkhalifah, T. (2022).
\newblock Target-oriented high-resolution elastic full-waveform inversion with
  an elastic redatuming method.
\newblock {\em Geophysics}, 87(5):R379--R389.

\bibitem[Lipari et~al., 2017]{Lipari2017}
Lipari, V., Urbano, D., Spadavecchia, E., Panizzardi, J., and Bienati, N.
  (2017).
\newblock Regularized tomographic inversion with geological constraints.
\newblock {\em Geophysical Prospecting}, 65(1):305--315.

\bibitem[Luo and Schuster, 1991]{LuoSchuster1991}
Luo, Y. and Schuster, G.~T. (1991).
\newblock Wave-equation traveltime inversion.
\newblock {\em Geophysics}, 56(5):645--653.

\bibitem[Madiba and McMechan, 2003]{Madiba2003}
Madiba, G.~B. and McMechan, G.~A. (2003).
\newblock Processing, inversion, and interpretation of a 2d seismic data set
  from the north viking graben, north sea.
\newblock {\em Geophysics}, 68(3):837--848.

\bibitem[Mosser et~al., 2020]{Mosser2020}
Mosser, L., Dubrule, O., and Blunt, M.~J. (2020).
\newblock Stochastic seismic waveform inversion using generative adversarial
  networks as a geological prior.
\newblock {\em Mathematical Geosciences}, 52:52--79.

\bibitem[Ravasi, 2025]{Ravasi2025_measurement_guided_diffusion}
Ravasi, M. (2025).
\newblock Geophysical inverse problems with measurement-guided diffusion
  models.
\newblock {\em arXiv}.

\bibitem[Saad et~al., 2024]{Saad2024}
Saad, O.~M., Harsuko, R., and Alkhalifah, T. (2024).
\newblock Siamesefwi: A deep learning network for enhanced full waveform
  inversion.
\newblock {\em Journal of Geophysical Research: Machine Learning and
  Computation}, 1(3):e2024JH000227.

\bibitem[Shepard, 1968]{Shepard1968}
Shepard, D. (1968).
\newblock A two-dimensional interpolation func- tion for irregularly-spaced
  data.
\newblock {\em Proceedings of the 23rd ACM National Conference, ACM}, page
  517–524.

\bibitem[Song et~al., 2021]{song2021ddim}
Song, J., Meng, C., and Ermon, S. (2021).
\newblock Denoising diffusion implicit models.
\newblock In {\em ICLR}.

\bibitem[Stork and Clayton, 1991]{Stork1991}
Stork, C. and Clayton, R.~W. (1991).
\newblock Migration velocity analysis.
\newblock {\em Geophysics}, 56(12):1961--1972.

\bibitem[Sun et~al., 2023a]{Sun2023}
Sun, J., Innanen, K., Zhang, T., and Trad, D. (2023a).
\newblock Implicit seismic full waveform inversion with deep neural
  representation.
\newblock {\em Journal of Geophysical Research: Solid Earth},
  128(3):e2022JB025964.

\bibitem[Sun et~al., 2023b]{Sun2023FWIreg}
Sun, P., Yang, F., Liang, H., and Ma, J. (2023b).
\newblock Full-waveform inversion using a learned regularization.
\newblock {\em IEEE Transactions on Geoscience and Remote Sensing}, 61:1--15.

\bibitem[Taner and Koehler, 1969]{Taner1969}
Taner, M.~T. and Koehler, F. (1969).
\newblock Velocity spectra-digital computer derivation and applications of
  velocity functions.
\newblock {\em Geophysics}, 34(6):859--881.

\bibitem[Tarantola, 1984]{Tarantola1984}
Tarantola, A. (1984).
\newblock Inversion of seismic reflection data in the acoustic approximation.
\newblock {\em Geophysics}, 49(8):1259--1266.

\bibitem[Tarantola, 1987]{Tarantola1987}
Tarantola, A. (1987).
\newblock {\em Inverse Problem Theory and Methods for Model Parameter
  Estimation}.
\newblock SIAM.

\bibitem[Taufik and Alkhalifah, 2024]{Taufik2024}
Taufik, M.~H. and Alkhalifah, T. (2024).
\newblock Wavenumber-aware diffusion sampling to regularize multiparameter
  elastic full waveform inversion.
\newblock {\em Geophysical Journal International}, 240(2):1215--1233.

\bibitem[Taufik et~al., 2024]{Taufik2024tomo}
Taufik, M.~H., Alkhalifah, T., and Waheed, U.~b. (2024).
\newblock Stable neural network-based traveltime tomography using
  hard-constrained measurements.
\newblock {\em Geophysics}, 89(6):U87--U99.

\bibitem[Van~Vliet and Verbeek, 1995]{vanvliet1995}
Van~Vliet, L.~J. and Verbeek, P.~W. (1995).
\newblock Estimators for orientation and anisotropy in digitized images.
\newblock In {\em ASCI}, volume~95, pages 16--18.

\bibitem[Virieux and Operto, 2009]{Virieux2009}
Virieux, J. and Operto, S. (2009).
\newblock An overview of full-waveform inversion in exploration geophysics.
\newblock {\em Geophysics}, 74(6):WCC1--WCC26.

\bibitem[Voronoi, 1908]{Voronoi1908}
Voronoi, G. (1908).
\newblock Nouvelles applications des paramètres continus à la théorie des
  formes quadratiques. premier mémoire. sur quelques propriétés des formes
  quadra- tiques positives parfaites.
\newblock {\em Journal für die reine und an- gewandte Mathematik},
  133:97–102.

\bibitem[Wang et~al., 2024]{Wang2024}
Wang, F., Huang, X., and Alkhalifah, T. (2024).
\newblock Controllable seismic velocity synthesis using generative diffusion
  models.
\newblock {\em Journal of Geophysical Research: Machine Learning and
  Computation}, 1(3):e2024JH000153.

\bibitem[Wang et~al., 2018]{Wang2018}
Wang, W., Yang, F., and Ma, J. (2018).
\newblock Velocity model building with a modified fully convolutional network.
\newblock 2018 SEG International Exposition and Annual
  Meeting:SEG--2018--2997566.

\bibitem[Wang et~al., 2004]{Wang2004}
Wang, Z., Bovik, A., Sheikh, H., and Simoncelli, E. (2004).
\newblock Image quality assessment: from error visibility to structural
  similarity.
\newblock {\em IEEE Transactions on Image Processing}, 13(4):600--612.

\bibitem[Woodward, 1992]{Woodward1992}
Woodward, M.~J. (1992).
\newblock Wave-equation tomography.
\newblock {\em Geophysics}, 57(1):15--26.

\bibitem[Wu and McMechan, 2019]{Wu2019}
Wu, Y. and McMechan, G.~A. (2019).
\newblock Parametric convolutional neural network-domain full-waveform
  inversion.
\newblock {\em Geophysics}, 84(6):R881--R896.

\bibitem[Yang and Ma, 2019]{Yang2019}
Yang, F. and Ma, J. (2019).
\newblock Deep-learning inversion: A next-generation seismic velocity model
  building method.
\newblock {\em Geophysics}, 84(4):R583--R599.

\bibitem[Yang and Ma, 2023]{Yang2023FWIreg}
Yang, F. and Ma, J. (2023).
\newblock Wasserstein distance-based full-waveform inversion with a regularizer
  powered by learned gradient.
\newblock {\em IEEE Transactions on Geoscience and Remote Sensing}, 61:1--13.

\bibitem[Zelt and Smith, 1992]{ZeltSmith1992}
Zelt, C.~A. and Smith, R.~B. (1992).
\newblock Seismic traveltime inversion for 2-d crustal velocity structure.
\newblock {\em Geophysical Journal International}, 108(1):16--34.

\end{thebibliography}

\end{document}